# A Gravity Model Analysis of Irish Merchandise Goods Exports under *Brexit**

Gerard Keogh

*Central Statistics Office, Dublin, Ireland*

*email: gerard.keogh@cso.ie

*Abstract:* This article uses the Gravity Model to explore the effects of *Brexit* on Irish merchandise goods exports. We embed our approach within the framework of the Structural Gravity Model that incorporates multilateral resistance terms to account for remoteness effects. Further we estimate the parameters of this model using the Poisson Pseudo Maximum Likelihood (PPML) method. Our data source is the Central Statistics Office's (CSO) 8-digit level CN (Combined Nomenclature) export data covering the years 1994 to 2016. We note that each of these features is novel in the context of modelling Irish exports.

Our key findings are a 1% increase in trade costs will result in a fall of 0.73% in goods exports value overall, a value considerably lower in magnitude than has been found previously but nonetheless much more in line with recent international research. A *Soft-Brexit* where the UK remains within the EU Customs Union will have no effect on the value of trade. Under a *Hard Brexit* where the UK leaves the Single Market and Customs Union and applies WTO tariffs to Irish goods exports, the value of goods exports to the existing EU-28 will fall by 1.4% while the overall value of goods exports will fall by €0.8bn on average. This overall fall in value will be borne mainly in the traditional sectors of agriculture, food and beverages, and textiles, where the hit will be between 9 to 12% in value terms. In the long term where Irish goods that had been sent to the UK are sent to EU countries instead and sold at EU country market prices, the effect of *Brexit* will result in a fall of €9.2bn in total export value. Many Irish export firms are unlikely to be able to sustain the associated fall in export sale values of between 50 and 95% and will likely cease trading. In national income terms we also estimate that GNI* will fall by up to 0.4% under WTO tariffs against the current baseline and by over 5%, if over the longer term Irish goods are diverted to other EU countries instead of the UK.

---

* The views expressed in this article are not necessarily those held by the CSO and are the personal responsibility of the author.

# 1. INTRODUCTION

In Ireland, our international trade policy strives to ensure our economic prosperity and well-being. Our trade policy is operationalised primarily through our membership of the EU. Crucially, for a small open economy like Ireland the advancement of trade under an international rules-based system is essential to sustaining economic growth and national income. However, this system is experiencing considerable change, whether it be the increasing economic and political influence of China, the impact of global financial crisis, the efforts by large economies to shape the global agenda in the G8 and G20, the rise of populism and its emphasis on trade protectionism and not least, for Ireland at any rate, *Brexit*, which is the focus of this paper. As a small state with limited resources and influence Ireland must navigate its way through these changes in the international trade regime. Crucial to this is a broad feel for trade relationships and a mature analysis of the evidence based on sound principles. An almost ubiquitous and valuable quantitative methodology available to those seeking to understand international trade relationships is the Gravity Model. The key features of the Gravity Model are that it assumes trade between two countries is a function of their economic masses and inversely of their distance apart – the resemblance to Newton's Law of Gravity is clear and hence the name. In this paper we use the Gravity Model as a platform to study the impact of *Brexit* on the value of Irish merchandise goods exports. Furthermore, we use this insight to consider how Ireland might respond so that the country's economic prosperity and our citizen's well-being are safeguarded.

For obvious reasons historically Ireland's main trading partner has been Great Britain (GB). We exported unprocessed agricultural produce, food and beverage goods while we imported coal, steel and consumer goods. Post-independence the reliance of the Irish economy on the agricultural sector and remittances, and the Government's adherence to a balanced budget left Ireland economically exposed. In the 1950's this situation became critical with Ireland suffering continuously high levels of emigration and unemployment while the rest of Europe was booming. Growth was non-existent and Ireland's trade balance falling continuously. Finally, in 1958 with the publication of *Economic Development* (1958) the country recognised its problems and set out a plan to deal with them via opening up the Irish economy to trade and foreign direct investment. In 1973 this process culminated in Ireland joining the EEC and even more significantly signing the Single European Act in 1987. The Single European Act liberalised trade in goods and services throughout the EU. In 1993 this gave rise to the twin system for monitoring merchandise trade, namely the Intrastat system for within EU movements and the SAD (Single Administrative Document) for goods outside the EU. Interestingly, the liberalisation of trade and services from 1993 onward could be said to be the starting point of the modern Irish economy, where exports to countries other than GB began to seriously take off largely as a consequence of foreign direct investment. In light of this our study of goods export movements covers the years from 1994 to 2016. We study these export movements through the lens of the Gravity Model to gain insight into the nature of trade relationships of modern Ireland. Intriguingly and somewhat surprisingly, both the nature of Irish goods exports which were largely final products in the past, but now include many intermediate products (with concomitant transfer pricing arrangements) and the main destination for these exports in value terms which turns out not to be Western Europe but the US, are outcomes that surely would have surprised the authors of *Economic Development* back in 1958.

This paper makes four significant and novel contributions to the analysis of merchandise exports based on the Gravity Model. The first is largely methodological. In contrast to previous studies of Irish trade data, such as Lawless (2010) we apply Poisson Pseudo Maximum Likelihood (PPML), see Santos–Silva & Teyrano (2006), to generate our model parameter estimates rather than the more conventional Ordinary Least Squares (OLS) method. PPML has been shown to produce less bias parameter estimates than OLS for the Gravity Model, and as a consequence has begun to supplant OLS as the estimation method of choice. We note that in line with previous Irish and international studies our analysis of aggregate goods exports gives an estimate of the elasticity of distance close to -1 for OLS. In contrast PPML gives substantially smaller effect for distance at about -0.7 that is consistent with the more recent international studies (see Santos–Silva & Teyrano, 2006). Meanwhile as a robustness check of dispersion under the Poisson model assumption we also estimate a Negative Binomial Pseudo Maximum Likelihood (NBPML) model. In this case we find the elasticity of distance is about -0.9, once again a smaller effect compared to the OLS value but closer to that value. Nonetheless a key novel finding of this study based on modelling aggregate data is the distance effect is smaller than that found previously for Irish goods exports.

The second novel contribution of this paper centres on the inclusion of so-called Multilateral Resistance Terms (MRT) in our Gravity Model. For example, multilateral resistances arise as a consequence of the preference of two neighbouring countries which are remote from Ireland to trade with each other rather than Ireland. Unfortunately the basic Gravity Model does not account for third-county remoteness and in some instances this creates a bias in the parameter estimates. Interestingly however Anderson and Van Wincoop (2003) deal with this problem by deriving a set of *structural gravity equations* that incorporate remoteness effects via multilateral resistance terms. In this paper we take the novel approach of including multilateral resistance effects with a view to ensuring our model is as theoretically sound as is practicable from an economic standpoint.

The third novel feature of this work is our primary data source is the Central Statistics Office's (CSO) unit level CN (Combined Nomenclature) goods export data covering the years 1994 to 2016. The CN code is an identifier describing the type of merchandise being traded and we use goods exports value data at the 8-digit CN level as our basic data in this analysis. To our knowledge detailed Irish exports data at this level of disaggregation has not been analysed elsewhere in the context of a Gravity Model study. To facilitate some comparisons with other studies we mention that we also conduct an initial analysis using aggregate goods exports values obtained as year by country totals of CN 8-digit export values.

Interestingly recently the CSO produced a report entitled *Bexit: Ireland and the UK in numbers* (CSO 2016a) which provides a comprehensive picture of all aspects of Ireland and UK interactions including trade flows. Meanwhile in 2015 the Economic and Social Research Institute (ESRI) also produced a comprehensive account of the impact of *Brexit* across the Irish economy entitled *Scoping the Possible Economic Implications of Brexit on Ireland* (ESRI 2015) and in their working paper series a report entitled *Modelling the Medium to Long Term*

*Potential Macroeconomic Impact of Brexit on Ireland* (ESRI 2016). Their analysis suggests the impact on trade flows could be 20% or higher in specific economic sectors with concomitant lower output in the Irish economy. Likewise in this context the Dept. of Finance (DoF, 2017) has conducted a comparison of the trade exposure of EU Member States to the UK in both goods and services. Interestingly their analysis looked at both overall and sectoral effects of *Brexit* using measures of size of trade exposures of EU countries to the UK. The results highlight that relative to other Member States Ireland is substantially more exposed in a number of goods sectors and this is particularly marked in the Agri-food sector. Accordingly this brings us to the final and most important novel contribution of this paper. Here we use our variant of the Gravity Model to study the effects *Brexit* at the overall and main NACE (first-digit) sector levels. Importantly our model is economically sensible as it is aligned with the Structural Gravity Model methodology which is derived from solid economic foundations (see Armington 1969 and Anderson 1979). So starting with a baseline model we compute Gravity Model parameter estimates across nine main economic sectors. Here our principle finding is a 1% increase in trade costs as measured by distance will result in a fall of 0.73% in goods exports value overall. With baseline parameter estimate computed we then further gauge the effect of *Brexit* on Irish goods exports under three counterfactual scenarios that describe *Soft, Hard* and *Long Term Substitution* forms of a *Brexit* respectively. At the overall state level our key findings are a *Soft-Brexit,* which broadly assumes the UK remains within the EU Customs Union, has no effect on goods trade, an outcome that is entirely expected. Under a *Hard Brexit* where the UK leaves the Customs Union goods exports will be impacted with a 1.4% drop on average with indigenous sectors of the economy taking the largest hit in value terms at 9 to 12%. Meanwhile based on a *Long Term Hard Brexit* goods export substitution scenario, trade costs would inevitably increase and goods exports value decrease by over €9bn on average and national income will fall by over 5% annually. Of course the brunt of this fall will be borne by the traditional sectors where goods exports to the UK will halve in the agriculture sector and be virtually annihilated in other traditional sectors such as food, beverage, textiles and wood products.

The remainder of this article is organised as follows. In section 2 we review the Gravity Model, discuss its appealing features, its flaws and what has been done to ameliorate them. We also briefly review previous Irish Gravity Model research. Section 3 looks at data trends in aggregate exports since the early nineties and their specific gravity features. Section 4 models the trade data using the Gravity Model and estimates this model using PPML. Section 5 discusses the impact of MRTs and builds a model that goes a long way to take account of these for Irish goods exports; this model is also estimated using PPML. Section 6 conducts the counterfactual analysis mentioned above while Section 7 concludes.

## 2. THE STRUCTURAL GRAVITY MODEL AND ITS ESTIMATION

### 2.1 Gravity Model features

The Gravity Model approach to analysing the value trade flows is based on the principle that trade between two countries is a function of trade costs. Trade costs are generally unobservable so proxies are used instead. Specifically, the gravity model for (international) trade has a great intuitive appeal as it relates trade between country pairs to their economic masses, usually taken as each country's GDP, and inversely to the distance between each country pair. Expressed in this way the Gravity Model within economics is simply an empirical tool for explaining trade flows between a country pair. The tool has a long history going back at least to Tinbergen (1962). The tool is also used to explain other types of international flows, most notably migration going back to Ravenstein (1885).

Based on this description the *cross-sectional* version of the Traditional Gravity Model for the value of trade flows from country *i* to country *j*, denoted by $T_{ij}$, is given by

$$T_{ij} = \alpha G_i^{\beta_1} G_j^{\beta_2} D_{ij}^{\beta_3} \eta_{ij} \qquad (1)$$

where $\alpha$ may be interpreted as world output and $\beta_1, \beta_2$ and $\beta_3$ are unknown constants to be estimated, $G_i, G_j$ are respective country GDPs, $D_{ij}$ is the distance between the country pair and $\eta_{ij}$ is a random disturbance forcing term. Broadly speaking distance is taken as a proxy for all factors that create trade resistance or trade costs (*e.g.* transport costs). Now, on the basis that the disturbance is independent and its expectation $E(\eta_{ij}) = 1$ we can take logs in (1) to get the *more usual* version which we call the Classical Gravity Model

$$\log(T_{ij}) = \beta_0 + \beta_1 \log(G_i) + \beta_2 \log(G_j) + \beta_3 \log(D_{ij}) + \varepsilon_{ij} \qquad (2)$$

with $\beta_0 = \log(\alpha)$, $\varepsilon_{ij} = \log(\eta_{ij})$ and $E(\varepsilon_{ij}) = 0$. In general estimating a Classical Gravity Model involves OLS estimation of the unknown parameters in (2) based on trade flow data for both exports and imports, generally aggregated at the country and year level. Given this is a log-linear model for trade flows, the coefficients $\beta_1, \beta_2$ and $\beta_3$ can be interpreted as the elasticities of trade with country GDP and distance respectively.

While Classical Gravity Model has been the basis for many studies (see Shepard, 2013, Ch 1) it has certain basic drawbacks, two of which we highlight. First, consider the impact on trade between country pair (*i, j*) of a change in trade costs between country pair (*i, k*). Clearly this is likely to impact the trade of country *j* but unfortunately the Classical Gravity Model cannot account for this purely economic flaw as it does not include a trilateral trade component. Anderson and Van Wincoop (2003), in their now famous paper *Gravity with Gravitas,* using a purely theoretical approach build a Structural Gravity Model which effectively deals with this kind of problem via the inclusion of Multilateral Resistance Terms (MRTs) that are correlated with trade costs. MRTs bear the intuitive interpretation that two countries will trade more with each other the more remote they are from the rest of the world. Clearly, because the Classical Gravity Model does not account for this remoteness there is a case of omitted variable bias in the model. Interestingly, Anderson and Van Wincoop (2003) also note the inclusion of MRTs has implications for the types of data that should be used in their Gravity Model. Specifically they say that trade flows should be expressed in nominal (current) prices and not in real (constant) prices. The reason for this is the MRTs occurring in their Structural Gravity Model are in fact price indices that enter the model as price deflators acting on remote country pairs. So, when a remote country pair, say Japan and South Korea trade together they are less inclined to trade with Ireland than would otherwise be the case. This additional or more correctly second order (resistance) effect on trade (distance being the first order effect) is directly captured by the MRT and therefore data analysed in real prices is deflated twice, with the real deflator being unsuitable for capturing second order effects. Moreover, Shepard (2013) emphasises this point stating "*deflating exports using different price indices, such as the CPI or the GDP deflator, does not adequately capture the unobserved multilateral resistance terms, and could produce misleading results*".

The second key weakness associated with the Traditional Gravity Model (1) is statistical. Two specific issues arise here and in another very influential paper, *The Log of Gravity*, Santos Silva & Tenreyro (2006) address both. First they show that OLS estimation of the parameters in (2) leads to biased estimates because the assumption of constant variance is incorrect (*e.g.* trade between country pairs tends to fall off with distance and so it is fair to say the associated disturbances will also tend to be smaller). In light of this Santos Silva & Tenreyro (2006) suggest re-expressing (1) using (2) giving what we call the PPML Gravity Model

$$T_{ij} = \exp(\beta_0 + \beta_1 \log(G_i) + \beta_2 \log(G_j) + \beta_3 \log(D_{ij})) \times \eta_{ij} \qquad (3)$$

In this form we can relax the constant variance assumption and for example assume $E(\eta_{ij}) \propto \text{Var}(\eta_{ij})$. This of course means that we have tacitly assumed country pair trade values $T_{ij}$ likely follow a Poisson (or more generally allowing for over/under dispersion a Negative Binomial distribution) rather than a lognormal distribution. When the parameters in (3) are estimated using conditional maximum likelihood under a Poisson model Santos Silva & Tenreyro (2006) call the procedure Poisson Pseudo Maximum Likelihood (PPML). In the more general case of over/under dispersion it is referred to here as Negative Binomial Pseudo Maximum Likelihood (NBPML). The second issue addressed by Santos Silva & Tenreyro (2006) is the treatment of zero trade flows; this has been explored in depth in recent gravity model estimations (see recent references in the attached Bibliography). Using (3) as a basis for analysis addresses this problem directly as actual rather than logged trade flow values appear on the LHS. This in turn reduces any bias of not including zero flows very considerably compared to OLS.

Of course the similarity of (3) with standard Poisson regression in a Generalised Linear Model (GLM) framework is obvious and so (3) may be readily estimated using standard GLM software with the log as canonical link function. In their work Santos Silva & Tenreyro (2006) demonstrate that PPML gives less biased estimates of Gravity Model parameters than OLS. As a consequence of this PPML has begun to supplant OLS as the tool of choice for Gravity Model estimation. A key innovation of this paper is that we apply PPML to Irish goods exports. Moreover, as robustness check on the assumed mean to variance relationship we also employ NBPML. We straightforwardly implement PPML and NBPML using the GLM software available in SAS and compute robust standard errors (SE) for PPML via clustering. Interestingly, we find initial evidence for over-dispersion suggesting the NBPML approach is more efficient. However, when we move to modelling the data at the CN-8 level the NBPML estimate of variance-covariance matrix (*i.e.* Hessian) turns out not to be positive-definite. As a consequence parameter estimates and their SEs are not reliable. Accordingly we focus our attention on PPML which provides more robust estimates at both the aggregated and disaggregated CN levels. Importantly, we believe this is the first instance where PPML (on NBPML) estimation has been applied in the context of a study specifically focussed on Irish goods exports. Also we mention the coefficients $β$ are, as before, the elasticities of trade with the respective continuous explanatory variable.

**2.2 Previous Gravity Model Research for Ireland**

Lawless (2010) provides the most recent and a good account of Gravity Model research for Ireland starting with Fitzpatrick (1984), accordingly it is not necessary to rehash those details here. Nonetheless it is import to note that all Gravity Model research for Ireland, including recent work has been based on the OLS estimation of the Intuitive Gravity Model. More recent work also mentioned by Lawless (2010) includes a study undertaken by Morgenroth (2009) on behalf of *InterTrade Ireland*. This study focussed on cross-border trade using sectoral data and found that the estimated trade flows were about 80% of the level predicted by the Intuitive Gravity Model. Meanwhile Lawless (2010) herself used the Intuitive Gravity Model on firm level export data to investigate whether the number of exporting firms and average export sales per firm are effected by trade costs. Consistent with the theory, she found all of the variables capturing language, internal geography, and import cost barriers have significant effects on the number of firms, but almost none of these variables have a significant relationship with the average export sales per firm. One other study worth mentioning is Walsh (2006), he examined whether the Intuitive Gravity Model could be applied services exports and found it fitted the data well in this case also.

**3. DATA SOURCES, IRISH EXPORTS TRENDS AND GRAVITY FEATURES**

Our primary data source is CSO's internal trade file at 8-digit CN level. This file has data on value and volume of goods export and import items by country, date and type (Intrastat/SAD). We select the exports and aggregate the export value as appropriate (*e.g.* total export value by year and destination country) to CN8 level. Our second data source is the CEPII Gravity Dataset (available at: http://www.cepii.fr/cepii/en/bdd_modele/bdd.asp)

– CEPII is the French Research Centre for International Economics. This file provides data on a number of variables and we extract a subset of these trade cost\effect variables which we list along with CSO variables in Table 1. We merge the CSO export trade data with CEPII dataset by country and year and base our subsequent analysis on this merged dataset. Of course not all variables in CEPII dataset are included in Table 1. For example, the contiguity variable is left out as Northern Ireland (NI) is the only contagious country and this means contiguity is perfectly collinear with our NI indicator variable. Moreover, to reflect the current status of NI and GB within the EU, their EU membership indicator is set equal to 1 and the NI and GB indicators are also set equal to 1. Later to assess the effect of *Brexit* we re-set the EU member indicator to 0 for both NI and GB. Meanwhile other CEPII variables are also dropped; these include GDP per capita, common ethnicity, common colony, date of independence and favoured trading status. Each of these variables is correlated to some degree with already included variables, GDP, country, population, distance, country or GATT membership, all of which are included variables. We introduce a time indicator variable to absorb time/panel effects present in our data in an effort to control for the movement in prices over time (see Shepard, 2013). Our analysis also requires us to compute multilateral resistance indices, in Section 5 we show how these are incorporated into our variation of the gravity model and for this purpose we use data on international bilateral trade flows taken from the Correlates of War database (see http://www.correlatesofwar.org/data-sets/bilateral-trade). Finally when we consider the *Hard Brexit* scenarios in Section 6 we have need of WTO tariffs, these have been sourced from: http://tariffdata.wto.org/Default.aspx? culture=en-US.

Table 1: Destination Country Trade Effect Variable List

| Variable Name | Scaling | Type |
| --- | --- | --- |
| GDP | log | Variable |
| Distance (km) | log | Variable |
| Area (km$^2$) | log | Variable |
| Population | log | Variable |
| Common Religion (Proportion Christian) | log | Variable |
| Time | log(Year-1992) | Variable |
| GB | - | Indicator |
| NI | - | Indicator |
| GATT/WTO member | - | Indicator |
| English is official or primary language | - | Indicator |
| EU member | - | Indicator |
| Euro Currency | - | Indicator |
| Common Legal System before Independence | - | Indicator |

To give a brief flavour of main features of Irish goods exports data on the merged file we have aggregated the goods export value by year and main destination area. Figure 1 shows the resulting trend in exports over the past 20 years. The basic pattern is clear, percentage export values to the EU and other areas have remained fairly stable over time. However, the key feature of the figure is the decline in the overall relative importance of GB & NI, dropping from around 25% in 1997 to roughly half that percentage in 2016 - in nominal terms the value has increased from €10bn in 1997 to about €14½bn in 2016. Meanwhile of course the relative importance of Canada & US has grown considerably from about 12% in 1997 to about double this figure in 2016. In nominal terms this growth has risen from about €5bn to in excess of €31bn over the period due largely to the importance of FDI in areas such as pharmaceuticals. The relative declining importance of the UK in value terms would tend to indicate the impact of *Brexit* on exports may be fairly manageable with the possible exception of the traditional sectors of agriculture and food. Of course this assumes the impact of extra trade costs for goods transiting across GB to the EU is not in play.

Figure 1: Irish Goods Exports by Year and Main Destination Area 1997-2016

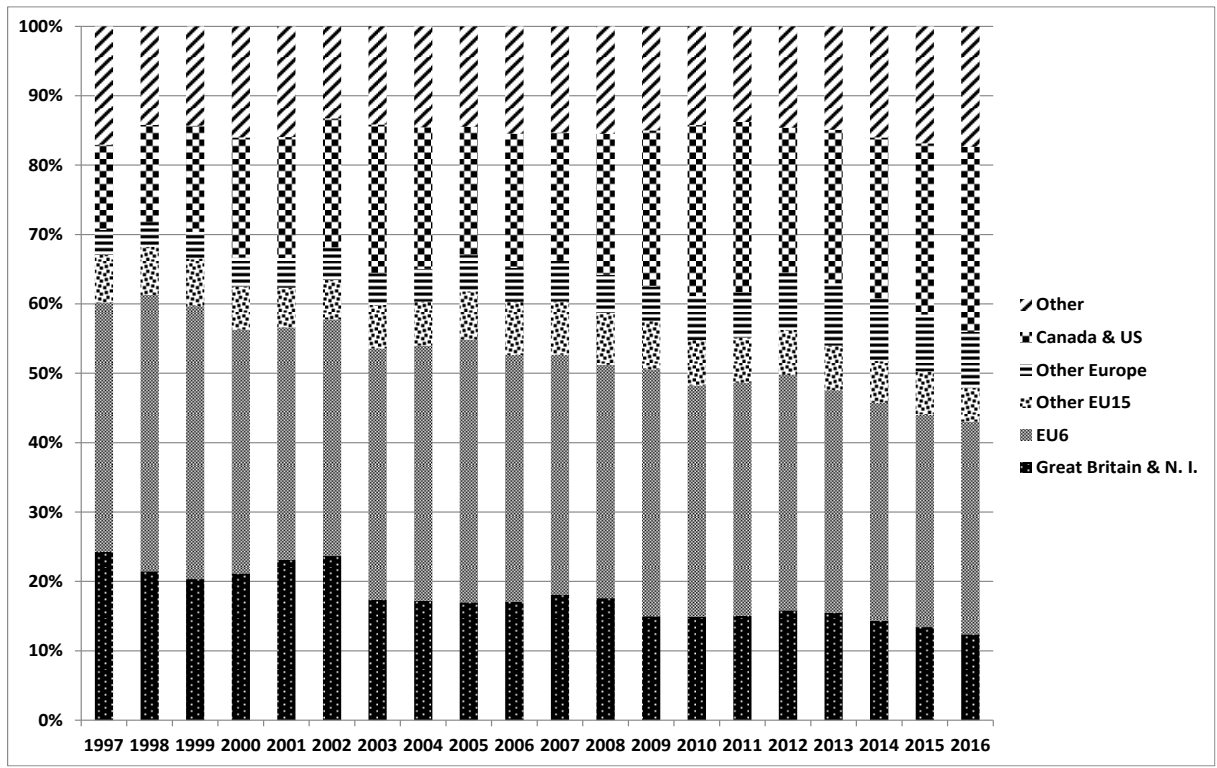

Looking at the data relationships more generally, Table 2 provides an analysis of the (Pearson) correlation between the variables in our datasets. All correlations are significant at the 1% level except for those few cells with a grey background. High correlations are evident between Export Value and GDP, (negative) Distance, Area, Population, GATT/WTO and EU membership. Accordingly, these variables should be important predictors for Export Value. Among the predictor variables GDP proves strongly correlated with population and area while area is simultaneously highly correlated with population, these relationships are to be expected as.

Table 2: Pearson Correlation Coefficients for Irish Goods Exports (all years)

| Variable Name | Export Value | GDP | Distance (km) | Time | Area (km²) | Population | Common Religion (Proportion Christian) | Great Britain | Northern Ireland | GATT/WTO member | English is official or primary language | EU member | Euro Currency | Common Legal System before Independence |
|---|---|---|---|---|---|---|---|---|---|---|---|---|---|---|
| **Export Value** | 1.00 | | | | | | | | | | | | | |
| GDP | 0.87 | 1.00 | | | | | | | | | | | | |
| Distance (km) | -0.43 | -0.32 | 1.00 | | | | | | | | | | | |
| Time | 0.06 | 0.18 | 0.00 | 1.00 | | | | | | | | | | |
| Area (km²) | 0.45 | 0.56 | -0.15 | 0.00 | 1.00 | | | | | | | | | |
| Population | 0.61 | 0.76 | -0.12 | 0.07 | 0.84 | 1.00 | | | | | | | | |
| Common Religion (Proportion Christian) | -0.08 | -0.06 | 0.07 | 0.00 | -0.19 | -0.18 | 1.00 | | | | | | | |
| Great Britain | 0.16 | 0.14 | -0.23 | 0.00 | 0.04 | 0.08 | 0.01 | 1.00 | | | | | | |
| Northern Ireland | 0.11 | 0.03 | -0.36 | 0.00 | -0.02 | -0.02 | 0.07 | 0.00 | 1.00 | | | | | |
| GATT/WTO member | 0.41 | 0.33 | -0.18 | 0.11 | 0.37 | 0.39 | 0.06 | 0.05 | 0.05 | 1.00 | | | | |
| English is official or primary language | -0.19 | -0.24 | 0.26 | 0.00 | -0.34 | -0.28 | 0.17 | 0.09 | 0.09 | -0.04 | 1.00 | | | |
| EU member | 0.46 | 0.36 | -0.61 | 0.08 | 0.07 | 0.12 | 0.11 | 0.21 | 0.21 | 0.24 | -0.14 | 1.00 | | |
| Euro Currency | 0.25 | 0.25 | -0.29 | 0.14 | -0.07 | -0.02 | 0.18 | -0.02 | -0.02 | 0.00 | -0.17 | 0.50 | 1.00 | |
| Common Legal System before Independence | -0.08 | -0.20 | 0.23 | 0.00 | -0.14 | -0.14 | -0.06 | 0.10 | 0.10 | 0.15 | 0.64 | -0.13 | -0.16 | 1.00 |

All correlations are significant at 1% level except those in grey cells

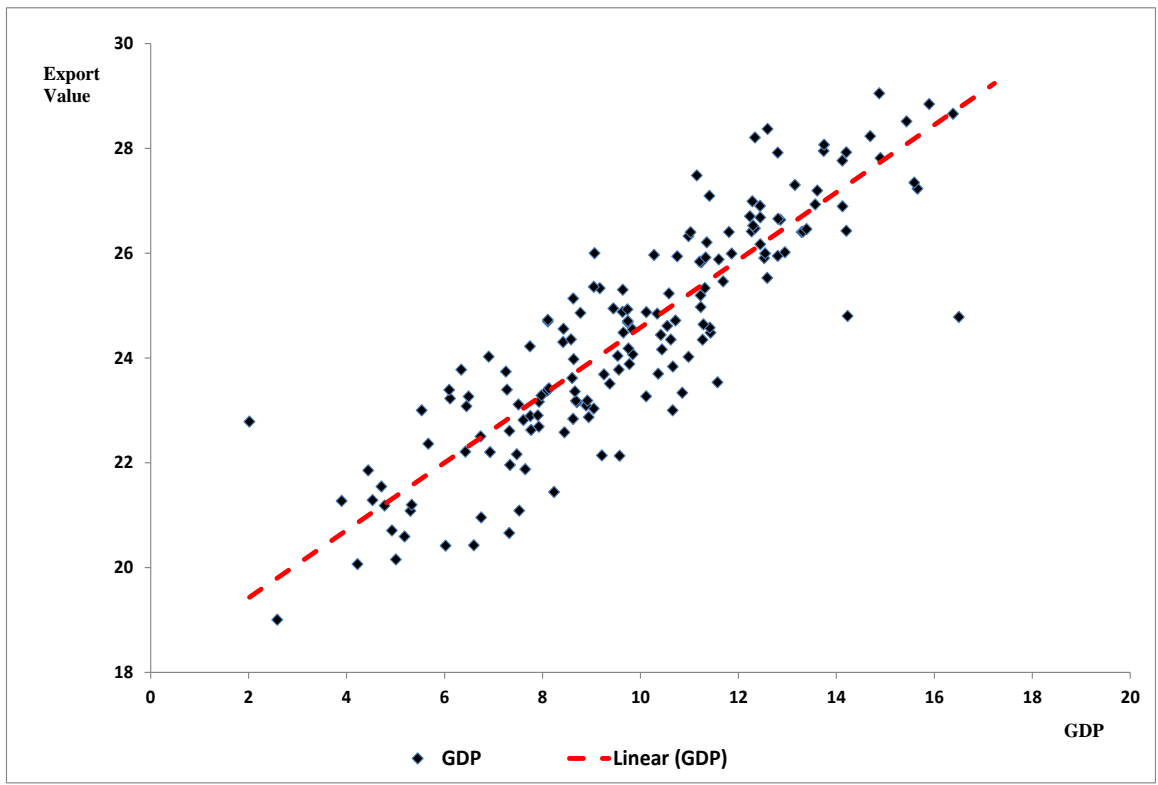

Figure 2: Relationship of Irish Goods Exports to Country GDP in 2016 (logs)

larger countries tend to have larger populations and GDP. In contrast GDP is negatively correlated with distance showing wealthier countries tend to be clustered together. Many of the other variables have relatively low correlation levels by comparison suggesting they will provide marginal value in modelling relationships. Nonetheless, the NI and GB indicator variables are of interest and correlations between these and other variables tends to be quite low. To some extent this is a mathematical artefact that arises because the computation of the denominator in the correlation is inflated when one of the variables is an indicator; a feature that in general tends to reduce the value of correlations based on indicator variables

Turning now to look at some key gravity model relationships in more detail, Figure 2 displays the relationship between goods exports value at the country aggregate level and destination country GDP for 2016 – all values are expressed on the log scale. The linear relationship in Figure 2 is clear and moreover the variation about the Linear (GDP) trend line looks constant. On this basis we can say that OLS applied to the Intuitive Gravity Model (2) will work well and provide an unbiased estimate of the elasticity coefficient of GDP.

Figure 3: Relationship of Irish Goods Export to Country Distance in 2016 (logs)

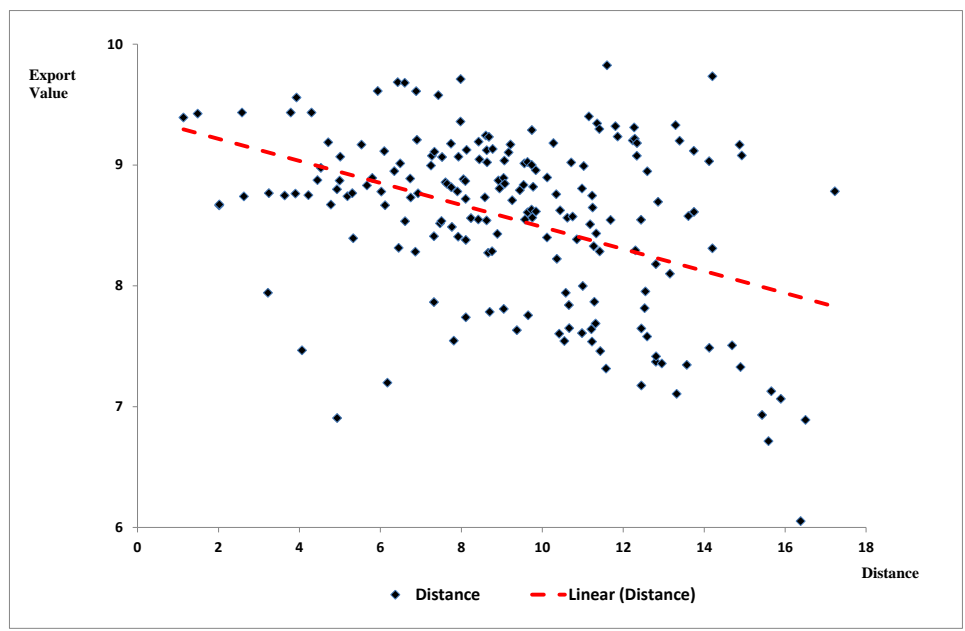

In Figure 3 the relationship between exports and country distance is displayed. As expected and predicted by theory a negative correlation with the Linear (Distance) is evident with the trend line falling with distance. Accordingly, working with the Intuitive Gravity Model (2) once again seems a good idea. However, it is also clear from the plot the variation of the data values about the trend line grows with distance, roughly doubling as the distance moves from about the value 1 through 15. Hetroskedasticity is therefore present and this would indicate that using Gravity Model (3) and estimating the parameters via PPML is the better option.

In Figure 4 we explore hetroskedasticity in a little more detail and plot the mean-variance relationship of the (log) goods export value for each year since 1994. The scatterplot shows a strong linear relationship between the mean and variance with the overlaid linear trend-line having a constant upward slope value of about 1.4. On this basis it is clear the mean and variance are not independent and therefore the constant variance assumption cannot be assumed. Accordingly OLS (even with robust SEs) is not an ideal method of parameter estimation. On the other hand it is clear there is a constant linear relationship between the mean and variance suggesting PPML or NBPML is likely much better suited to parameter estimation and should give unbiased estimates. Moreover, on the basis the dispersion evident in Figure 4 comes in at about 1.4, and not 1 as assumed in PPML, it would seem NBPML is preferable.

Figure 4: Mean-Variance Relationship of Irish Goods Exports log(Value) by Year

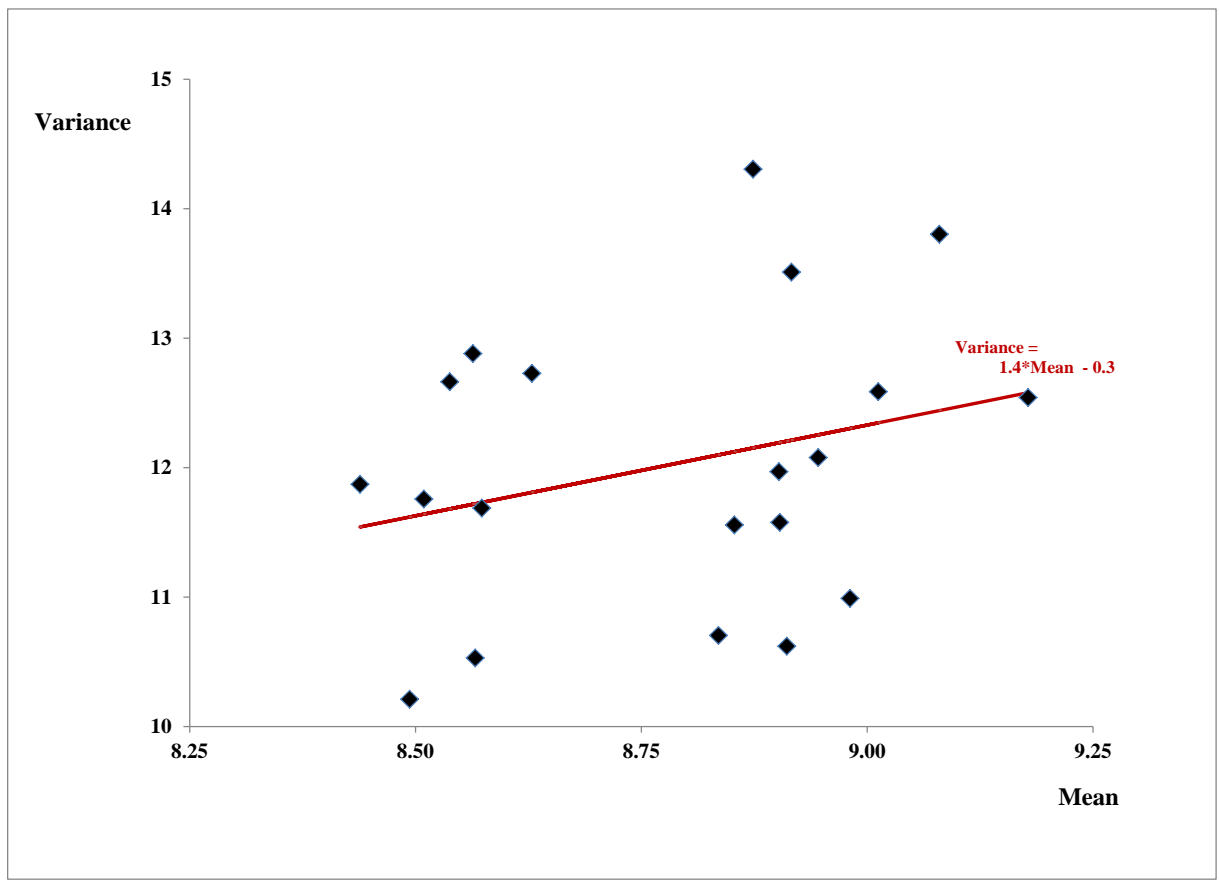

.

## 4. INITIAL MODELLING AND RESULTS

Our starting point for modelling the data is goods exports value aggregated by year and destination country, this gives us a panel of 4,061 non-missing observations on which to base our first set of estimates. We use OLS to generate parameter estimates for each trade volume effect variable listed in Table 1 associated with Irish goods exports at time $t$ using the Exports Classical Gravity Model

$$\log(X_{j,t}) = \beta_0 + \beta_1 \log(G_{j,t}) + \beta_2 \log(D_{j,t}) + \beta_3 \log(P_{j,t}) + \beta_4 \log(A_{j,t}) + \beta_5 \log(t_j) + \sum_{k=7} \gamma_k I_{jk} + \varepsilon_{j,t} \quad (4)$$

with goods exports from Ireland to country $j$ labelled $X_j$ and $I_{jk}$ is the $k^{th}$ indicator variable given in Table 1. Correspondingly, we use PPML to generate parameter estimates from the Exports PPML Gravity Model

Table 3: Parameter Estimates for Aggregated Irish Goods Exports by Gravity Model Type

| Variable Name | Classical Gravity Model OLS Estimate | | | | Natural Gravity Model PPML Estimate | | | | Natural Gravity Model NB Estimate | | | |
|---|---|---|---|---|---|---|---|---|---|---|---|---|
| | Parameter Estimate OLS | % effect | $C_v$ | Signif. Level | Parameter Estimate PPML | % effect | $C_v$ | Signif. Level | Parameter Estimate NB | % effect | $C_v$ | Signif. Level |
| Intercept | -8.84 | | 0.06 | * | -15.40 | | 0.14 | * | -6.42 | | 0.12 | * |
| GDP | 1.16 | | 0.01 | * | 1.41 | | 0.06 | * | 1.07 | | 0.02 | * |
| Distance (km) | -0.99 | | 0.04 | * | -0.62 | | 0.10 | * | -0.88 | | 0.06 | * |
| Time | -0.42 | | 0.07 | * | -0.20 | | 0.26 | * | -0.33 | | 0.11 | * |
| Area (km²) | -0.08 | | 0.15 | * | -0.12 | | 0.15 | * | -0.12 | | 0.19 | * |
| Population | 0.01 | | 2.36 | 0.72 | -0.77 | | 0.17 | * | 0.01 | | 4.91 | 0.84 |
| Common Religion (Proportion Christian) | -0.04 | | 0.23 | * | 0.17 | | 0.15 | * | 0.00 | | 4.17 | 0.81 |
| Great Britain | -1.28 | -72 | 0.22 | * | -0.32 | -27 | 0.62 | 0.11 | -1.05 | -65 | 0.21 | * |
| Northern Ireland | -0.70 | -50 | 0.43 | * | -1.58 | -79 | 0.15 | * | -0.88 | -59 | 0.33 | * |
| GATT/WTO member | 0.50 | 65 | 0.11 | * | 0.64 | 90 | 0.30 | * | 0.31 | 37 | 0.24 | * |
| English is official or primary language | 0.63 | 88 | 0.11 | * | 0.19 | 21 | 0.72 | 0.16 | 0.27 | 31 | 0.51 | 0.05 |
| EU member | 0.13 | 13 | 0.80 | 0.11 | -0.24 | -22 | 0.38 | * | 0.15 | 16 | 0.84 | 0.23 |
| Euro Currency | 0.28 | 32 | 0.41 | * | 0.50 | 65 | 0.23 | * | 1.49 | 345 | 0.16 | * |
| Common Legal System before Independence | 0.10 | 10 | 0.70 | 0.16 | 0.89 | 145 | 0.16 | * | 0.30 | 35 | 0.45 | 0.03 |
| * => Significant at 1% level | $R^2 = 0.85$ | | | | pseudo-$R^2$ Poisson = 0.88 | | | | pseudo-$R^2$ NB = 0.20 | | | |

$$X_{j,t} = \exp(\beta_0 + \beta_1 \log(G_{j,t}) + \beta_2 \log(D_{j,t}) + \beta_3 \log(P_{j,t}) + \beta_4 \log(A_{j,t}) + \beta_5 \log(t_j) + \sum_{k=7} \gamma_k I_{jk}) \times \eta_{j,t}$$

(5)

while the Exports NBPML Gravity Model is also estimated via (5) but with an alternative non-canonical log inverse link function (see McCullagh and Nelder, 1989).

The results of model estimation are given in Table 3 where in all cases the coefficient of variation $C_V = \frac{\sigma}{|\mu|}$ (*i.e.* SE of the parameter estimate divided by the parameter estimate) is computed based on robust SEs and a significance level of 1% is reported with a *. For the trade effect variables listed in Table 1 we can interpret the parameter estimate for each continuous trade effect variable directly as the elasticity of trade *w.r.t.* relative changes in that variable. For the $I^{th}$ indicator the percentage effect change is given by

$$(\exp(\hat{\beta}_I) - 1) \times 100\% \tag{6}$$

Meanwhile for the OLS model we also report the model fit (adjusted) $R^2$ value in Table 2 and for GLM models a pseudo-$R^2$ statistic (see Heinzl, and Mittlbock, 2003) given by

$$R^2_{Model\ Type} = 1 - \left(\frac{D(\hat{\mu})}{D(\bar{y})}\right) \tag{7}$$

is reported -.here $D(\bar{y})$ is the deviance in the Null (intercept only) Model and $D(\hat{\mu})$ is the deviance of the fitted model. Importantly, we caution these deviance measures are not directly comparable to the $R^2$ as deviance estimates are not based on minimising the sums of squares in a GLM.

The results in Table 3 are intriguing. With the exception of the population variable all three methods give parameter estimates with similar levels of precision as the corresponding $C_V's$ turn out to be similar. As a consequence of this all three methods tend to give a similar set of variables that are significant at the 1% level. Of course there are some differences with the population variable being the most notable (see next page). Looking at the overall measures of fit both the OLS and PPML models provide very high levels of fit but the NB model is poor in comparison suggesting estimates from this model should be ignored. We note this verdict is reinforced by the size of the log-likelihoods where smaller is better, for the PPML model this is -5.3x10[8] while the NB model (which allows for over-dispersion) gives much higher value -4.1x10[4]. In light of this we conclude there is little evidence the apparent over-dispersion in Figure 4 affects the model estimates leading us to conclude the PPML model is robust for these data. Accordingly our analysis proceeds based on the OLS and PPML models.

Looking now at GDP (recall data are in logs) the OLS parameter estimate (*i.e.* elasticity) is 1.16 while PPML is 1.41. As expected based on theory and in line with the very high positive correlation value given in Table 2, both estimates are positive with a 1% increase in GDP across destination countries inducing a 1.41% increase in goods exports value in Ireland according to PPML Focussing on the trade effect parameter estimate associated with distance. The OLS elasticity of goods exports with distance is as expected negatively correlated with value -0.99, this value of course is similar to that found in previous Irish studies based on OLS (*e.g.* Lawless 2010) and elsewhere more generally (see Head & Meyer, 2013, §4). In contrast to OLS the PPML estimate is considerably smaller in absolute magnitude at -0.62 and is much closer to the correlation figure -0.43 reported in Table 2. The PPML estimate is also much closer to the -0.7 value reported for country pairs studied with PPML in Santos Silva & Tenreyro (2006). Thus, here as with other studies (e.g. Santos Silva & Tenreyro 2006 or Piermartini & Yotov, 2016), PPML suggests that distance has a substantially smaller effect on Irish goods exports as has previously been found – this is a new finding for Irish goods exports that cannot be overemphasised. Indeed this finding applies irrespective of the changing nature of Irish goods exports over recent times. Importantly, a likely explanation for this smaller negative elasticity value is that Ireland is an island nation, so the marginal cost of say exporting the same goods to Britain as opposed to say France may not be too great. Moreover, this finding also suggests that increases in transit costs associated with moving Irish goods to the EU via Britain after *Brexit,* while still relative high, may not be as prohibitive as might be initially suspected.

Looking at the other variables in Table 2 we see firstly that time is significant with an elasticity of -0.2 with PPML, this suggests the importance of merchandise exports over time is falling in relative terms. Interestingly, the area of the destination country has a negative effect on trade, thus Ireland tends to export less to larger countries proportionately under these models. The population parameter is also negative under PPML with a value -0.77 showing a 1% increase in population will reduce goods exports by 0.77%. At first sight this seems strange, but it is a common feature of PPML models incorporating a population variable (see Flaherty 2015). Essentially, since GDP per capita must be positive, accordingly population must be a divisor in the model resulting in a negative coefficient on the log scale. However, these estimated parameter values do contradict the correlations given in Table 2 where goods exports were positively correlated with both area and population while area and population pair also has a positive correlation of 0.84. So-called *sign reversal or confounding* is at play here - this arises when correlation estimates are contrasted with multiple regression parameter estimates when there is also correlation among predictors. This situation likewise occurs with the common religion variable which has a PPML parameter estimate of 0.17 Here the PPML model seems to give the more sensible outcome as most of Ireland's main trading partners are Christian or might be thought of as having been 'Christian'. Accordingly, a 1% increase in the Christian population of a trading partner will result in a 0.17% increase in Irish goods exports to that destination country. It is also worth pointing out that the change in sign of the parameter for religion and population under OLS as compared to PPML at first sight seems worrying, however Piermartini & Yotov (2016, p34) point out this feature often occurs where OLS and PPML estimates are compared.

The GB indicator provides us with contradictory picture regarding the relative importance of GB as an export destination. While a negative estimate is observed reflecting the declining importance of GB as an export destination, there is a substantial difference between OLS = -1.28 and PPML = -0.32 estimates, with the latter not statistically significant at 1%. On the basis of this analysis GB exports are at a percentage level similar to the mean percent level of most other countries – to a degree this is a comforting observation with *Brexit* looming as it shows GB no longer dominates Irish exports. Meanwhile, controlling for all other factors and using (6) the NI indicator shows a 79% lower percent level of exports *w.r.t* to the global average. On this basis and from a size, proximity and declining importance (see Figure 1) perspective, NI therefore does not seem to be too vital an export market for the rest of Ireland. Accordingly the *Brexit* of NI shouldn't adversely impact on the rest of Ireland except possibly within specific traditional sectors – this issue will explored in more detail in the Sections 5 and 6 of this paper.

As might be expected, being a member of GATT/WTO shows a strong relationship with a trade effect of 90% under PPML and 65% under OLS. Naturally the large positive effect of GATT/WTO will balance somewhat the large level drops observed for GB and NI. Also under PPML having English as a common language is not significant. On the basis that PPML is the more correct model specification, this shows that English as common language is far less important than might be thought. Indeed this probably just reflects the fact that many countries already use English as an international language for communication and so having it as a common language shouldn't be overly important. Looking at EU membership we again see a sign reversal between OLS and PPML. The latter indicator shows a 22% fall which seems counterintuitive. However, when considered in light of the growing importance of the USA as an export destination the finding is less surprising. Moreover, this negative relationship also agrees with the findings of Morganroth, (2008, see Table 2 Single European Act Indicator) who suggests the probable cause is a falloff in intra-EU trade and an increase in external trade due to globalisation. Meanwhile the percentage effect associated with being a member of the Euro currency is 65% under PPML showing being in the Euro area enhances Ireland's trade overall, this value is higher but nonetheless is broadly in line with the 30% effect found by Baldwin (2006). Mind you it must be said that Santos Silva and Tenreyro (2010) found virtually no effects on trade for the Euro, after taking into account the high level of trade integration of Eurozone members even before they formed a common currency. Finally, the results for 'Common Legal System before Independence' show that countries which share a British type legal system are strong export destinations with a 145% trade effect overall. Once again this will tend to balance the large GB and NI drop in percentage goods export levels.

For completeness, we also note that more a detailed analysis of parameter estimates for model equation (5) by main (NACE) industry sector under the PPML Model is provided in Table 1A in the Appendix; we call these estimates the Basic Model estimates. This analysis is based on goods exports trade values at the detailed CN8 level. This analysis puts more flesh and bones on the aggregate analysis given in Table 2. Our NACE product grouping is derived from the Eurostat Correspondence between CN and Prodcom classifications with the first 4-digits of Prodcom in turn matching the NACE Rev. 2 Industry Classification. Of course creating this correspondence is known to be a difficult task and the result can be poor in some CN areas but we only use the first 2-digits of Prodcom/NACE and here the mapping is fairly sound. With our 2-digit NACE code in place we further aggregate to the nine groups shown in Appendix Table 1A. We mention that at this level parameter estimates remain stable as there are plenty of observations within each group; below this level we have found the $R^2$ values tend to fall considerably. On the basis that PPML is the more correct model formulation we provide Basic Model estimates for PPML only in Table 1A with all starred parameter estimates significant at the 1% level. We also note that all analysis and results from here on will be based on PPML estimation of the CN8 data cross classified by main (NACE) industry sector.

## 5. MULTILARERAL RESISTANCE MODELLING AND RESULTS

As noted above a key flaw in the Traditional Gravity Model is that it fails to take account of remoteness MRT effects. Failure to take account of MRTs has been referred to as the *gold medal mistake* of traditional gravity model analysis by Baldwin and Taglioni (2007). One of the solutions Feenestra (2004) suggests for handling multilateral resistance involves augmenting the Traditional Gravity Model (1) with exporter and importer fixed effects

$$T_{ij} = \alpha G_i^{\beta_1} G_j^{\beta_2} D_{ij}^{\beta_3} e^{\theta_i d_i + \theta_j d_j} \eta_{ij} \tag{8}$$

Here $\theta_i$ and $\theta_j$ are respectively the outward and inward multilateral resistance parameters to be estimated and $d_i$ and $d_j$ are dummies identifying corresponding exporter and importer fixed effects - we note for panel data Oliveria & Yatov (2012) augment (8) with exporter-time and importer-time fixed effects. Unfortunately (8) only applies in the case where the data are a cross-section (or more generally a panel ) of bi-directional flows and so is unsuitable for a single or unidirectional export country analysis of the type under study here. However, a work around that goes some way to account for MRTs is available (see Yotov et. al. 2016) based on remoteness indices constructed as output (for exporting countries) and expenditure (for importer countries) weighted averages of bilateral distance. For each exporting country in the world these exporter/output remoteness weight indices at time *t*, ( *i.e.* year) are given by

$$r_{j,t} = \sum_k \left( D_{jk,t} \times \frac{E_{k,t}}{Y_t} \right) \tag{9}$$

where $E_{k,t}$ is expenditure by the importing country in the relevant year on all imports and $Y_t$ is the total world exports in that year. We compute these yearly remoteness indices using bilateral total trade data from the Correlates of War database and use these to augment the Exports PPML Gravity Model (8) with exporter effects. For completeness we also add Irish importer country fixed effects labelled $\chi_j$ in an effort to absorb importer country MRTs, giving our full Remote Exports PPML Gravity Model for panel data

$$X_{j,t} = \exp(\beta_0 + \beta_1 \log(G_{j,t}) + \beta_2 \log(D_{j,t}) + \beta_3 \log(P_{j,t}) + \beta_4 \log(A_{j,t})$$
$$+ \sum_t \pi_t \log(r_t) + \sum_{k=1}^{7} \gamma_k I_k + \sum_{j=1}^{} \chi_j I_j) \times \eta_{j,t} \qquad (10)$$

Along with the basic variables included in (8), equation (9) has 23 export remoteness time effect indices and 190 country fixed effect dummies. Of course, as we now have a set of remoteness (MRT) indices covering all time periods, the continuous time variable included previously is no longer necessary as the remoteness indices now absorb time effects. We apply this model to our CN8 exports data and present the results by main industry sector in Table 4. For comparison with counterfactual scenarios in the next section we refer to the parameter estimates in Table 4 as *Baseline* estimates.

Table 4: *Baseline* Parameter Estimates by NACE Sector Group for Irish Goods Exports using PPML with MRTs

| | Baseline Remoteness Gravity Model PPML Parameter Estimates | | | | | | | | |
|---|---|---|---|---|---|---|---|---|---|
| | Main Industry Sector | | | | | | | | |
| Variable Name | Agriculture, Forestry & Fishing | Mining & Quarrying | Food & Beverage | Textiles | Wood & Paper | Chemicals, Pharma & Rubber | Metals & Machinery | Other Products | All Sectors |
| Intercept | -4.799 | 4.319 | -6.024* | 30.811* | 7.011* | -6.027* | -19.811* | 31.506* | -11.248* |
| GDP | 0.432 | 0.658* | 0.595* | 0.208 | 0.567* | 1.265* | 1.190* | 0.058 | 1.065* |
| Distance (km) | -0.943* | -0.765 | -0.088 | -1.452* | -0.476* | -1.822* | -0.703* | -1.079* | -0.730* |
| Area (km²) | 0.274* | 0.892* | 0.163* | -0.202* | 0.006 | -0.310* | -0.078* | -0.091 | -0.009 |
| Population | 0.515 | -1.239 | -0.099* | 0.553* | -0.384* | 0.183* | -0.151* | -0.066 | -0.34* |
| Common Religion (Proportion Christian) | 0.065 | -0.064 | 0.018 | -0.175* | -0.056 | 0.111* | -0.016 | 0.071 | 0.197* |
| Great Britain | 2.058* | 2.726 | 2.188* | -0.592 | 2.773* | -3.943* | -1.235* | 4.661* | -0.608 |
| Northern Ireland | 3.758* | 0.622 | 2.538* | -1.646 | 1.726* | -4.782* | -1.233* | 3.012* | -1.528* |
| GATT/WTO member | 1.201 | 0.583 | -0.567* | 1.146* | -0.536* | 0.068 | -0.025 | -1.819* | -1.047* |
| English is official or primary language | -1.508* | 0.263 | -0.347* | 0.347 | 0.867* | 0.407 | 1.623* | 0.836* | 1.067* |
| EU member | 0.975* | 0.019 | 0.979* | 0.100 | -0.155 | -0.502* | 0.452* | 0.362 | -0.193 |
| Euro Currency | 0.513* | 1.66* | -0.144 | 0.519* | 0.147 | -0.210 | -0.263* | 1.113* | -0.016 |
| Common Legal System before Independence | 0.792 | -1.133 | 0.660* | 0.197 | -0.024 | 1.132* | -0.393 | -1.129* | 0.074 |
| Export Remoteness (1%) (percentage of (5%) indicies significant) | 0 0 | 100 " | 0 " | 100 " | 100 " | 0 44 | 100 " | 100 " | 66 " |

* denotes significance at 1% level
Pseudo $R^2$: 0.89

First, to gauge the effect of handling MRTs in the Baseline Remoteness Model (10), we can compare the parameter estimates in the All Sectors column of Table 4 with the corresponding PPML parameter estimates in Table 3. Comparing GDP which is a continuous variable, so the parameter is the elasticity, the remoteness indices and country effects reduce the elasticity from 1.41 to 1.065, a drop of almost 25%. Meanwhile the effect on distance and therefore on direct trade costs is an increase in absolute magnitude from -0.62 to -0.73, nearly 18%. Furthermore, the effect on most of the other variables is also substantial with the three last variables, EU Member, Euro currency and Common Legal System going from being statistically significant to not being so under Model (10). The conclusion here is clear, neglecting remoteness results in biased estimates and here the effect is generally in excess of 20% (in absolute value) which is quite substantial. On the basis of this and studies elsewhere (see Yotov et. al. 2016) as well as the theoretical foundations of the Structural Gravity Model of which (10) is a unidirectional adaptation, we speculate the results in Table 4 are among the most credible based on gravity modelling of Irish goods exports values.

In Table 4 the overall the elasticity of GDP *w.r.t* goods exports at 1.065 shows us that a 1% increase in GDP across Ireland's goods export destination countries will return almost a 1.07% increase in goods exports value. Within sectors goods export elasticities in the traditional sectors (1st five columns) are all positive and substantially less than one. In Agriculture for example the estimate is 0.432 so a 1% increase in GDP across all export destinations will result in a 0.43% increase in goods export values. In general exports in the traditional sectors are likely to have low unit values so a small percentage value response to an increase in GDP is to be expected. Equally these sectors will suffer least from a general fall in our GDP which is comforting with *Brexit* on the horizon. The Chemicals & Pharma sector has an elasticity of 1.265 while Metals & Machinery is 1.19. Thus a 1% increase in GDP across goods export destinations will be associated with substantial increases of 1.27% and 1.19% respectively in Irish goods exports. Once again it is likely these sectors produce high unit value goods so substantial increases in goods export value responses are to be expected in a general global expansion.

Looking at the key variable Distance which reflects direct transports cost associated with traded goods the overall elasticity is -0.73, accordingly a 1% increase associated with the distance Irish goods exports travel will mean the value of those exports will drop by 0.73% - this is quite a large distance effect. Across the sectors the distance parameter estimate is, as expected, negatively correlated with goods exports. Agriculture exports will fall by 0.94% for 1% increase in distance. Specifically, a simple example suggests agricultural produce exported from Ireland to France at 1,557km instead of Britain at 494km (distances by road taken from https://www.freemaptools.com), a distance factor of 3.15, will trigger the value of those exports to fall by a factor of 2.96 - with *Brexit* an impact of this size will mean goods export values fall to 34% ((1/2.96)*100%) of their current levels over time – this is of course is a substantial drop in a sector which is important to the indigenous Irish economy. Among the other sectors distance has the greatest effect in Chemicals & Pharma where the elasticity is -1.822. Interestingly without MRTs the corresponding value from Appendix Table 1A is -0.419 which gives the impression that an increase in distance doesn't overly affect export values, as might be expected given the trade in these goods is global so they already travel substantial distances. However, controlling for remoteness, (the tendency of distant countries to trade more with each other) via MRTs shows that distance and therefore trade cost have quite a severe impact on goods export values in the sector. Given the importance of multinational companies in this sector the relevance of this finding should not be ignored by those involved in promoting these industries.

The estimates in Table 4 show that importer country area has a small effect on goods export values. The same applies to population size with the exception of Mining & Quarrying where population size is negatively correlated with export values and has an elasticity of -1.239. This suggests goods exports in this sector tend to be to smaller countries that are also likely quite close. The remaining continuous variable Common Religion also has a very small effect on goods exports; indeed here most of the sectoral parameter estimates are not significant reinforcing the relative lack of importance of a shared religion in determining exports – of course this is not particularly surprising.

Turning to the indicator variables the percent effect of GB and NI respectively using (6) are -46% and -78%. Note this is a level effect and therefore simply reflects the degree of importance of these two UK zones in terms of Ireland's export values compared to the global average level. Indeed this is also a reflection of the declining importance of the UK to Ireland's goods exports, particularly in value terms. Nonetheless across the sectors the picture is more varied. In Agriculture the parameter estimates are 2.058 and 3.758 for GB and NI respectively; these values give percent effects of 683% and 4,186% respectively. Clearly these huge positive level effects serve only to demonstrate the relative importance of the UK to Irish agricultural exports. With the exception of Textiles the UK also tends to be vital to the other traditional sectors though not to as great an extent as with agriculture. Meanwhile, at the other extreme the relative unimportance of the UK as an export destination to the Chemical & Pharma sector is shown with very large negative elasticity of -3.943 and -4.782 for GB and NI respectively.

Overall the parameter estimate for the GATT/WTO indicator is negative which is a little surprising given that most countries are members of these organisations. Across the sectors five parameter estimates are not significant. Meanwhile destination countries using English are also shown to have a positive effect on goods exports, but intriguingly within agriculture English is associated with relatively lower levels of goods exports. As noted above parameter estimates for EU member, EU Currency and Common Legal System are not significant. However, as might be expected the parameter estimate the Agriculture is significant for EU membership and currency, the same applies for Metals & Machinery. For Chemical & Pharma though we can see that EU membership is -0.502 equating to a -39.5% level effect relative to the global level. While this might seems a little surprising the Chemical & Pharma exports to EU countries other than one destination country tend not to be overly substantial when we have controlled for GB.

The last row of Table 4 shows the percentage of remoteness indices that were found to be significant. It is clear where this value is 100% the inclusion of MRTs is important but for agriculture and food & beverage exports it is clear MRT are not relevant. Given goods in these sectors tend to be traded with near neighbours this observation seems reasonable. Indeed comparing the result in Table 4 with those in Appendix Table 1A where no remoteness indices are used, the effect of MRTs in agriculture is quite small while in textiles the parameter estimates can be very different and therefore biased where MRTs are ignored.

To summarise the findings, it is clear that including MRTs has a substantial impact on parameter estimates. Equation (10) is a variation of the Structural Gravity Model which has a sound economic basis unlike classical gravity models which are empirically based. Accordingly the results in Table 4 are much more credible than those in Table 3 (or indeed Table 1A) where the estimates are likely biased. Thus the remoteness model provides a more sound representation of the nature of Irish goods exports. In terms of trade costs the key finding of this section is the substantial effect of distance, specifically a 1% increase associated with the distance Irish exports travel will mean the value of goods exports will drop by 0.73%. Meanwhile in agriculture exports shifted to France in the event of *Brexit* could see export values fall by up to 66% over time.

## 6. *BREXIT* COUNTERFACTUAL ANALYSIS

Broadly speaking there are at least two approaches to counterfactual 'what if' analysis of international trade. The first and largely dominant approach in the international trade literature is econometric analysis which usually takes place within the context of a general equilibrium model. Indeed more often than not this is based on the Structural Gravity Model of Anderson and Van Wincoop (2003) which in turn is derived from CES (Constant Elasticity of Substitution)-Armington Model (see Armington 1969). Typically there are three stages or levels in the Structural Gravity Model: (i) Direct or Partial Equilibrium (PE) effects; (ii) Conditional General Equilibrium (CGE) effects; and (iii) Full Endowment General Equilibrium (FE) effects – see Larch & Yotov (2016), p16 for equations and further details relating to the Structural Gravity Model. The key advantage of this approach is that it gives a full econometric picture of the impact of a counterfactual.

The second approach is statistical or data based. Here the analyst simulates the effects on trade of different conditions such as GB not being in the EU and compares the result with baseline estimates. In principle this approach is relatively straightforward but care needs to be taken to ensure the simulation actually mirrors reality. In this section we will explore econometric PE effects directly by making changes to indicator variables and indirectly via the statistical approach to counterfactual analysis, with a view to gauging the effect *Brexit* may have on our goods exports and its impact on our economy more generally.

In this section we consider three counterfactual scenarios that might describe different forms of *Brexit* as it applies to merchandise exports from Ireland; these are:

- *Soft Brexit*: For merchandise exports this means the UK leaves the EU but stays fully within the EU Customs Union and so does not apply any tariff to goods originating in the EU, it also applies EU tariffs to goods from third countries.
- *Hard Brexit*: A *Hard Brexit* for merchandise exports means the UK leaves the EU and leaves Customs Union. It applies ad-valorem WTO tariffs to Irish goods and sets its own standards of compliance for goods it imports from the EU. As a consequence a hard border is created on the Island of Ireland.
- *Long Term Hard Brexit*: As above, a *Hard Brexit* for merchandise exports where the UK applies WTO ad-valorem tariffs to Irish goods but Irish exporters respond over time by diverting those goods to other EU markets with the values/prices being modified by features of the new EU destination country.

For completeness we provide an additional table in the appendix Table 2A, this offers insight into the impact on Irish goods exports of *Regulatory Alignment,* this is where the UK leaves the EU but NI stays in the Customs Union while GB leaves. GB then applies ad-valorem WTO tariffs to Irish goods (and indeed goods from NI). Basically this is a half-way house between a *Soft* and *Hard Brexit* and the figures in Table 2A bear this out with NI estimates remaining equal to their *Soft Brexit* values while the GB estimates equal their *Hard Brexit* Values.

Now, within the econometric approach PE effects are the initial and generally the strongest direct response effects on our exports/imports to a change in trading conditions, these effects only are explored in this study. First baseline estimates are established and then the counterfactual changes are made and the model parameters re-estimated. In our case the baseline estimates are given by the PPML estimates shown in Table 4. The first counterfactual we explore is *Soft Brexit*. Here we simply set the EU indicator variable to zero for both GB and NI and otherwise assume all trading conditions remain the same. On this basis we proceed to re-compute the PPML counterfactual model parameter estimates; these are given in Table 5 where they are referred to as *Soft-Brexit* estimates.

Comparing *Soft Brexit* estimates in Table 5 to those that apply to goods export values under the existing arrangements in Table 4, it is clear that the both tables are almost identical. This is appealing as distance and most of the other variables have not changed and so, trade costs associated with, for example, distance can be expected remain the same. The GB and NI indicators have changed but of course this is nothing other than an indicator level shift due to the values of the combination of EU and GB indicators or EU and NI indicators respectively being altered from (1, 1) to (0, 1). Thus, for example, the combination of the GB indicator parameter estimate for All Sectors in Table 4 is -0.608 while the EU indicator parameter estimate is -0.193. In Table 5 the EU value stays at -0.193 but the GB value changes to -0.800 = -0.608-0.193 due to the EU indicator value now being set to zero while the GB indicator is kept at one. Accordingly, Table 5 simply reflects the pure indicator value change associated with GB and NI being removed from the EU assuming all other trading conditions remains the same. Therefore, under this Remoteness Gravity Model a *Soft Brexit* has absolutely no effect on trading conditions, an outcome that is entirely expected.

Table 5: *Soft Brexit* Parameter Estimates by NACE Sector Group for Irish Goods Exports using PPML with MRTs

| | Soft Brexit Remoteness Gravity Model PPML Parameter Estimates | | | | | | | | |
|---|---|---|---|---|---|---|---|---|---|
| | Main Industry Sector | | | | | | | | |
| Variable Name | Agriculture, Forestry & Fishing | Mining & Quarrying | Food & Beverage | Textiles | Wood & Paper | Chemicals, Pharma & Rubber | Metals & Machinery | Other Products | All Sectors |
| Intercept | -4.799 | 4.319 | -6.024* | 30.811* | 7.011* | -6.027* | -19.811* | 31.506* | -11.248* |
| GDP | 0.432 | 0.658* | 0.595* | 0.208 | 0.567* | 1.265* | 1.19* | 0.058 | 1.065* |
| Distance (km) | -0.943* | -0.765 | -0.088 | -1.452* | -0.476* | -1.822* | -0.703* | 1.079* | -0.730* |
| Area (km²) | 0.274* | 0.892* | 0.163* | -0.202* | 0.006 | -0.31* | -0.078* | -0.091 | -0.009 |
| Population | 0.515 | -1.239* | -0.099* | 0.553* | -0.384* | 0.183* | -0.151* | -0.066 | -0.340* |
| Common Religion (Proportion Christian) | 0.065 | -0.064 | 0.018 | -0.175* | -0.056 | 0.111* | -0.016 | 0.071 | 0.197* |
| Great Britain | 3.033* | 2.746 | 3.167* | -0.492 | 2.619* | -4.444* | -0.784* | 5.023* | -0.800* |
| Northern Ireland | 4.733* | 0.641 | 3.517* | -1.546 | 1.571* | -5.284* | -0.781* | 3.374* | -1.721* |
| GATT/WTO member | 1.201 | 0.583 | -0.567* | 1.146* | -0.536* | 0.068 | -0.025 | -1.819* | -1.047* |
| English is official or primary language | -1.508* | 0.263 | -0.347* | 0.347 | 0.867* | 0.407 | 1.623* | 0.836* | 1.067* |
| EU member | 0.975* | 0.019 | 0.979* | 0.100 | -0.155 | -0.502* | 0.452* | 0.362 | -0.193 |
| Euro Currency | 0.513* | 1.66* | -0.144 | 0.519* | 0.147 | -0.210 | -0.263* | 1.113* | -0.016 |
| Common Legal System before Independence | 0.792 | -1.133* | 0.660* | 0.197 | -0.024 | 1.132* | -0.393 | -1.129* | 0.074 |
| Export Remoteness (1%) (percentage of (5%) indicies significant) | 0 0 | 100 " | 0 " | 100 " | 100 " | 0 44 | 100 " | 100 " | 66 " |

* denotes significance at 1% level
Pseudo R² : 0.89

The second counterfactual scenario we consider extends the *Soft Brexit* set up but involves the imposition of WTO tariffs by the UK on Irish goods exports - a set up that is referred to as a *Hard Brexit*. This scenario places a hard border on the island of Ireland. We consider ad-valorem tariffs only taken from the WTO Tariff Database for year 2017. As noted in the InterTrade Ireland Report (2017) these tariffs are set on goods classified according to the 6-digit Harmonised System (HS6) and can vary from 0 to 80%. Importantly, HS6 is identical to the CN trade classification system at the six digit level. Accordingly we match our CN8 data to the WTO HS6 at the 6-digit correspondence level and apply the WTO tariffs at this 6-digit level to Irish goods exported to GB. We assume the price of the good in the UK remains the same but the imposition of the tariff reduces the value to Irish exporters by the tariff rate. This reduced value is taken as the export value for goods exported to the UK in our PPML estimation of the Remoteness Gravity Model. The resulting parameter estimates are given in Table 6.

Table 6: *Hard Brexit* Parameter Estimates by NACE Grouping for Irish Goods Exports using PPML with MRTs

| | Hard Brexit Remoteness Gravity Model PPML Parameter Estimates | | | | | | | | |
|---|---|---|---|---|---|---|---|---|---|
| | Main Industry Sector | | | | | | | | |
| Variable Name | Agriculture, Forestry & Fishing | Mining & Quarrying | Food & Beverage | Textiles | Wood & Paper | Chemicals, Pharma & Rubber | Metals & Machinery | Other Products | All Sectors |
| Intercept | -4.720 | 4.326 | -5.947 | 30.909* | 6.849* | -6.049* | -19.815* | 31.521* | -11.324* |
| GDP | 0.433 | 0.658* | 0.592* | 0.208 | 0.569* | 1.265* | 1.19* | 0.058 | 1.066* |
| Distance (km) | -0.941* | -0.765 | -0.091 | -1.453* | -0.478* | -1.822* | -0.702* | 1.079* | -0.731* |
| Area (km$^2$) | 0.271* | 0.892* | 0.165* | -0.203* | 0.008 | -0.310* | -0.078* | -0.091 | -0.009 |
| Population | 0.516 | -1.239* | -0.098* | 0.555* | -0.387* | 0.183* | -0.151* | -0.066 | -0.341* |
| Common Religion (Proportion Christian) | 0.064 | -0.064 | 0.018 | -0.175* | -0.056 | 0.111* | -0.015 | 0.071 | 0.197* |
| Great Britain | 2.919* | 2.744 | 3.030* | -0.602 | 2.593* | -4.470* | -0.796* | 5.012* | -0.850* |
| Northern Ireland | 4.643* | 0.635 | 3.403* | -1.657 | 1.553* | -5.318* | -0.823* | 3.364* | -1.787* |
| GATT/WTO member | 1.179 | 0.583 | -0.561* | 1.144* | -0.535* | 0.068 | -0.025 | -1.819* | -1.048* |
| English is official or primary language | -1.505* | 0.263 | -0.344* | 0.345 | 0.875* | 0.407 | 1.623* | 0.836* | 1.068* |
| EU member | 0.974* | 0.019 | 0.979* | 0.101 | -0.156 | -0.500* | 0.452* | 0.362 | -0.191 |
| Euro Currency | 0.522* | 1.660* | -0.143 | 0.513* | 0.153 | -0.212 | -0.264* | 1.112* | -0.019 |
| Common Legal System before Independence | 0.796 | -1.133* | 0.661* | 0.198 | -0.027 | 1.132* | -0.393 | -1.130* | 0.073 |
| Export Remoteness (1%) (percentage of (5%) indicies significant) | 0  0 | 100  " | 0  " | 100  " | 100  " | 0  39 | 100  " | 100  " | 66  " |

* denotes significance at 1% level
Pseudo R$^2$ : 0.90

We compare the *Hard Brexit* parameter estimates in Table 6 to their corresponding *Soft Brexit* estimates in Table 5 and not the *Baseline* estimates in Table 4 as the latter pair are identical except for the level shift effect mentioned above. It is clear that the both tables are once again almost identical. It is somewhat surprising that the impact of the tariffs on trade cost as measured by distance is negligible. This is especially so since, as noted above, some WTO tariffs are set at 80%. However, their impact is also negligible as in general high tariffs in excess of 50% tend to apply to a small number of products mainly in the Food & Beverage sector and so reduce the overall export value by only a few million Euro. That said, the comparison of the estimates in Tables 5 and 6 shows that goods exported to the UK will not experience a trade cost effect under the imposition of a WTO ad-valorem tariff – subject to frictionless movement of Irish goods into the UK this would be very comforting outcome for Irish exporters. Of course in a *Hard Brexit* the UK remains outside the EU and here the parameter estimates of GB and NI in Tables 5 and 6 reflect this effect. For GB the All Sectors estimate drops from -0.800 to -0.850, using (6) the percentage level equivalents of these amounts are -55.1% and -57.3% respectively. Of course for indicator variables we can use a variant of (6) to compute the overall relative change in the percentage value of Irish exports according to

$$\exp(\hat{\beta}_{I,Hard} - \hat{\beta}_{I,Soft}) - 1 ) \times 100\% \qquad (11)$$

while for continuous variables *Hard vs. Soft* elasticities can be directly compared, so the relative change is simply

$$( (\hat{\beta}_{I,Hard} - \hat{\beta}_{I,Soft}) - 1 ) \times 100\% \qquad (12)$$

For statistically significant key predictors these quantities are reported in Table 7 (in grey scale). Thus for the GB indicator the relative effect of a *Hard Brexit* is a drop of 4.9% in goods export value. In the NI case we find the fall overall is 6.5%. If, in addition we consider a worst case two SE situation, the relative effect for GB would result in a drop 6.8% while overall for NI trade the effect would be a drop by 9.4%. We note this value is very similar to the value of 9% given in the InterTrade Ireland Report (2016) which it is important to note includes ad-valorem and other non-tariff trade levies and barriers. Interestingly, on the bottom row of Table 7 we also quantify the impact of a *Hard Brexit* on the overall value of Irish exports to the existing EU-28 assuming the UK has left the EU. To be precise, the overall impact is arrived at by adding the tariff adjusted value of exports to the UK to the EU-27 value under the *Hard* (or below *Long Term Hard*) scenario and then computing the percent change in this value for the resulting EU-28 relative to the EU-28 *Soft Brexit* value. We note this is an exact calculation that is available to us because we apply the tariff directly to the detailed CN6/HS6 export values classified by country. On foot of this the overall percent impact on the value of Irish goods exports to the EU-28 is expected to fall by just 1.4%. This is also a key novel finding of this study.

Table 7: Percent relative impact of a *Hard Brexit* on key statistically significant parameter estimates

|  | Main Industry Sector | | | | | | | | |
|---|---|---|---|---|---|---|---|---|---|
| Variable Name | Agriculture, Forestry & Fishing | Mining & Quarrying | Food & Beverage | Textiles | Wood & Paper | Chemicals, Pharma & Rubber | Metals & Machinery | Other Products | All Sectors |
| GDP | 0.2 | 0.0 | -0.5 | 0.0 | 0.4 | 0.0 | 0.0 | 0.0 | 0.1 |
| Distance (km) | -0.2 | 0.0 | 3.4 | 0.1 | 0.4 | 0.0 | -0.1 | 0.0 | 0.1 |
| Great Britain | -10.8 | -0.1 | -12.8 | 22.4 | -2.6 | -2.6 | -1.2 | -1.1 | -4.9 |
| Northern Ireland | -8.6 | -0.9 | -10.8 | 7.2 | -1.8 | -3.3 | -4.1 | -1.0 | -6.4 |
| Impact of *Hard Brexit* on Overall Value of Irish Exports to EU 28 (i.e. including UK) compared to a *Soft Brexit* | | | | | | | | | |
| Impact (%) | -6.4 | 0.0 | -6.8 | -5.6 | -1.6 | -0.5 | -0.5 | -0.4 | -1.4 |

Now, returning to Table 6 a varied picture emerges across the sectors based on GB and NI indicators. First, other than for Textiles all parameter estimates for the traditional sectors are positive, showing the levels of trade with the UK in these sectors are above the global average. In Agriculture the GB parameter estimates are $\hat{\beta} = 3.033$ in Table 5 and $\hat{\beta} = 2.919$ in Table 6. Using (6) the respective percent equivalents are 1,975% and 1,752%. In terms of the relative impact of a *Hard Brexit* on the value of Irish Agricultural exports the figure in Table 7 shows an expected fall of 10.8% compared to a *Soft Brexit*. Meanwhile with a *Hard Brexit* the corresponding effective drop in the value of cross-border agricultural export trade turns out to be 8.6% on average. More interestingly, whether it be a *Soft* or *Hard Brexit* scenario, a worst case two SE situation could see agriculture exports to GB drop in value by up to 17.6% while for NI it would be 15.1%, an effect that is considerably lower than the 66% initial estimate identified in Section 5. Similar effects will be felt by exporters of Food & Beverages products to GB with expected falls in the relative value of exports being 12.8% while for NI the average drop will be 10.8%. Moreover a worst case two SE could see the fall in the relative value of trade being 16.4% for GB and 14.4% for NI. Meanwhile at the overall EU-28 level there is about a -6% impact on export values across the traditional sectors of the Irish economy on foot of a *Hard Brexit*. More generally the figures show the UK effect of about a 10% drop in goods exports values translates into an overall EU-28 drop of about 5 to 6%. Looking at the non-traditional sectors in Table 6 the Chemicals & Pharma sector has negative parameter estimates indicating percent value levels are below global averages. However the GB parameter changes very little so there will be no effect here. As ad-valorem tariffs are virtually zero across this sector this result is not surprising. Meanwhile from Table 7 the relative drop in goods exports value of Metals & Machinery trade with GB will be 1.2% while for NI it will be 4.1%.

Our third counterfactual scenario builds further on the second and looks at what the effects on Irish exports will be if Irish exporters try to divert as much of their UK exports to EU countries. We call this scenario the *Long Term Hard Brexit* as it might reflect the exports landscape if, for example, the UK were to impose transit or other non-tariff barriers to Irish goods entering the UK and then Irish exporters had to find new EU markets for those goods. This indeed is a realistic probability since, as has been shown by the Dept Of Finance (DoF 2017), Ireland has a revealed comparative advantage in many indigenous sectors. Our approach here is relatively straightforward, within each year and for each group of CN8 goods exported to the UK we look for a corresponding CN8 substitute within the EU. Where a substitute is available we change the destination country to the alternative EU country and adjust the value to the corresponding EU value. Where an alternative EU destination does not exist we apply the WTO tariff regime as previously described and leave the UK as the destination country. With the dataset reconfigured in this way we re-run our PPML model estimates with the results shown in Table 8. Two key points are important to highlight here, first we have assumed perfect instantaneous substitution of goods across markets and this occurs without affecting the price in the destination. A nice additional feature therefore of our approach is that it captures the instantaneous effect of a tariff or other trade policy. Second, we have assumed transit to the new destination will not be via the UK so no transit tariffs apply and therefore all goods export costs remain a function of distance and other predictors in the model.

Table 8: *Long Term Hard Brexit* Parameter Estimates by NACE Grouping for Irish Exports using PPML with MRTs

| | Long Term Hard Brexit based on Simulated Substitutes Remoteness Gravity Model PPML Parameter Estimates | | | | | | | | |
|---|---|---|---|---|---|---|---|---|---|
| | Main Industry Sector | | | | | | | | |
| Variable Name | Agriculture, Forestry & Fishing | Mining & Quarrying | Food & Beverage | Textiles | Wood & Paper | Chemicals, Pharma & Rubber | Metals & Machinery | Other Products | All Sectors |
| Intercept | -5.463 | 4.382 | -4.831* | 32.216* | 7.735* | -6.828* | -17.840* | 32.484* | -11.236* |
| GDP | 0.380 | 0.645* | 0.566* | 0.223 | 0.510* | 1.281* | 1.152* | 0.055 | 1.065* |
| Distance (km) | -0.918* | -0.783 | -0.146* | -1.784* | -0.604* | -1.840* | -0.713* | 1.083* | -0.746* |
| Area (km$^2$) | 0.303* | 0.891* | 0.157* | -0.241* | 0.029 | -0.313* | -0.102* | -0.1 | -0.015 |
| Population | 0.552 | -1.219* | -0.051 | 0.707* | -0.291* | 0.194* | -0.093 | -0.061 | -0.318* |
| Common Religion (Proportion Christian) | 0.069 | -0.060 | 0.031* | -0.135* | 0.044 | 0.108* | 0.012 | 0.073 | 0.206* |
| Great Britain | 1.715* | 0.710 | 0.047 | -4.069* | 0.862* | -7.891* | -3.766* | 0.752 | -3.772* |
| Northern Ireland | 4.014* | -0.085 | 1.469* | -4.405* | 0.384 | -6.607* | -2.396* | 0.858 | -3.182* |
| GATT/WTO member | 1.319 | 0.650 | -0.524* | 1.082* | -0.425* | 0.084 | -0.066 | -1.839* | -1.055* |
| English is official or primary language | -1.385* | 0.284 | -0.403* | 0.141 | 0.594* | 0.413 | 1.457* | 0.795* | 1.020* |
| EU member | 1.194* | 0.033 | 1.034* | 0.371* | 0.166 | -0.393* | 0.535* | 0.389 | -0.080 |
| Euro Currency | 0.684* | 1.658* | -0.040 | 0.54* | -0.122 | -0.239 | -0.314* | 1.030* | -0.057 |
| Common Legal System before Independence | 0.762 | -1.152* | 0.699* | 0.383 | 0.016 | 1.100* | -0.296 | -1.150* | 0.086 |
| Export Remoteness (1%) (percentage of (5%) indicies significant) | 0 0 | 100 " | 0 " | 100 " | 100 " | 0 83 | 100 " | 100 " | 66 " |
| * denotes significance at 1% level Pseudo R$^2$ : 0.89 | | | | | | | | | |

We compare the *Long Term Hard Brexit* results in Table 8 to the *Soft Brexit* estimates of Table 5 to gauge the full impact Brexit could have on our goods exports if trading conditions with the UK deteriorated. Unlike the straightforward *Hard Brexit* case discussed above it is clear that parameter estimates have changed. The All Sector distance value has moved from -0.730 to -0.746. We can compare these using (11) to compute the relative impact on the elasticity of distance to exports value by simply substituting *Long Term Hard Brexit* for *Hard Brexit* in formula. The resulting values for statistically GDP and Distance estimates across sectors are given in Table 9 (in grey scale). Overall for distance we can see there is a 2.2% increase in trade cost associated with finding new markets for exports. However, contrary to expectations the parameter estimate associated agricultural produce increases from -0.941 under a *Soft Brexit* to -0.918 in the *Long Term* scenario. Strikingly this will result in a 2.7% fall in trade costs that are a function of distance. Counterintuitively this suggests the overall price of agricultural products exported to the EU is slightly lower than their UK substitutes. Meanwhile for Textiles the trade cost effect is 22.9% and for Wood & Paper it is 26.9% while the remaining sectors are only marginally affected. Clearly these results show that trade costs associated with finding alternative EU markets for goods currently exported to the UK will weigh heavily on the traditional sectors of Textiles and Wood & Paper.

Table 9: Percent relative impact of a Long Term *Hard Brexit* on key statistically significant parameter estimates

| Variable Name | Main Industry Sector | | | | | | | | |
|---|---|---|---|---|---|---|---|---|---|
| | Agriculture, Forestry & Fishing | Mining & Quarrying | Food & Beverage | Textiles | Wood & Paper | Chemicals, Pharma & Rubber | Metals & Machinery | Other Products | All Sectors |
| GDP | -12.0 | -2.0 | -4.9 | 7.2 | -10.1 | 1.3 | -3.2 | -5.2 | 0.0 |
| Distance (km) | -2.7 | 2.4 | 65.9 | 22.9 | 26.9 | 1.0 | 1.4 | 0.4 | 2.2 |
| Great Britain | -73.2 | -86.9 | -95.6 | -97.2 | -82.7 | -96.8 | -94.9 | -98.6 | -94.9 |
| Northern Ireland | -51.3 | -51.6 | -87.1 | -94.3 | -69.5 | -73.4 | -80.1 | -91.9 | -76.8 |
| Impact of *Long Term Hard Brexit* on Overall Value of Irish Exports to EU 28 (i.e. including UK) compared to a *Soft Brexit* | | | | | | | | | |
| Impact (%) | -35.8 | -12.6 | -42.2 | -33.8 | -46.1 | -10.7 | -20.7 | -27.9 | -19.2 |

Now, looking at the GB and NI indicator effects the overall values change from -0.800 to -3.772 for GB and -1.721 to -3.182 for NI. Clearly these are substantial movements that in percent terms equate to a level shift from -57.3% to -97.7% for GB and -83.3% to -95.8% for NI. In Table 9 this is reinterpreted as a relative impact using (10) where we see the effect is a drop of 94.9% for GB and 76.8% for NI respectively – values much more in line with the 66% drop initially identified in Section 5. As in Table 7 we can also quantify the overall EU-28 level impact on Irish goods exports of, in this case, a *Long Term Hard Brexit*. Here this measure is central as it bundles together the effect of the tariff on those remaining goods sent to the UK with the value of all goods diverted from the UK to the EU (now priced at the appropriate EU country value), along with those originally sent to the EU. Accordingly the real value derived from the exported goods across the existing EU-28 is measured and this is compared to *the Soft Brexit* value. Thus, in Table 9 the overall effect of *Brexit* where Irish exporters divert goods currently sent to the UK to new EU markets equates to a 19.2% drop in goods export value. This finding is vitally important since Irish agricultural exports to the UK are highly exposed to any change in UK trade policy post *Brexit* – a policy that is likely to favour imports from non-EU countries in sectors where UK has a comparative disadvantage (*e.g.* cheap agricultural goods) (see Dept. of Finance, DoF 2017, p18). In this context our finding carries even more weight because we control for MRT effects, thus the 19.2% drop is not biased by preferences among non-EU states remote from the UK to trade among themselves. Clearly, what the analysis in this paragraph tells us is that if price of Irish goods exported to the UK stays as is, then in the long term the value of goods exports to the UK will become negligible while the value of cross-border exports would run at about one-quarter of their current levels. For all practical purposes this would kill off the Irish goods export trade to the UK. Moreover, the extra cost of diverting those exports to other EU countries will reduce the overall value of Irish goods exports by close to 20% across the whole EU-28. Accordingly, in tariff equivalent terms Irish exporters should stay in UK markets after *Brexit* where tariffs stay well below 20%, above this they will likely cease trading unless their export price can be reduced by at least that percentage amount.

For the agricultural sector the GB indicator parameter estimate value changes from 3.033 in Table 5 to 1.715 in Table 8 – using (6) the corresponding percent levels are 1,752% and 456%. In Table 9 we can see this translates into a relative drop in the level of exports of 73.2% on average. However a portion of this drop would be compensated by a rise in these exports to the EU where the parameter estimates change from 0.975 to 1.194 equating to a relative growth of 25% (using (10)). For NI the relative impact on the Food & Beverage sector is -87.1% which represents a virtual wipe out in the value of Food & Beverage exports to NI. A similar effect occurs for exports to GB. Meanwhile, the distance or trade cost associated with these exports is 65.9% in Table 9, so if the new destination for existing Food & Beverage exports to the UK is to be the EU, then Irish exporters will have to absorb this substantial trade cost effect. Comparison of export values at the EU-28 level shows a severe drop of 42.2% in export values overall when UK tariffs and transport costs to the EU are taken into account. Clearly a large drop in Irish producer prices will be necessary to offset the increased costs and ensure the export price stays competitive. It seems likely this would make trading in the sector very difficult and in the long term could lead to many companies having to cease trading in Ireland.

As with other traditional sectors of the economy Table 9 also shows a very substantial relative impact on Textiles for GB and NI with falls of 97% and 94% respectively in the value of exports. This however is compensated by the EU indicator increase of 24.3% (using (10)), reflecting percent levels change from 10.6% to 44.9%. Finding new EU markets that could absorb an additional 70% of textile imports is likely to be challenging without a minimum drop in the export price of at least 33.8% based on overall EU-28 percent impact on the export value. Meanwhile looking at the effects on the Chemicals & Pharma sector Table 9 shows relative impacts of 96.8% and 73.4% for GB and NI respectively. Nonetheless, the trade cost impact as measured by distance in this sector is negligible, a reflection of the fact that these goods have very small tariffs and high unit values so the marginal cost of diverting to the EU is likely small.

Table 10: Effect of Brexit Tariff scenarios on GNI*

|  |  | Value/ Estimate | Actual Change on Soft Brexit Export Value | Adjusted GNI* Estimate | Change on GNI* Value |
|---|---|---|---|---|---|
|  |  | €bn | €bn | €bn | % |
| GNI* Value |  | 181.0 | - | - | - |
| Export Value Estimates | Soft Brexit (no change) | 115.5 | - | - | - |
|  | Hard Brexit | 114.7 | 0.8 | 180.2 | -0.4 |
|  | Long Term Hard Brexit | 106.3 | 9.2 | 171.8 | -5.1 |

Of course the analysis so far provides an interesting insight based on data over a long time horizon. Also of interest, for comparison with GDP say, are more recent trends. Accordingly we focus on the average value of goods exports in 2015 and 2016 with a view to gauging the impact of *Brexit* on Ireland's overall national income under each of the three counterfactual scenarios considered here. Importantly we cannot compute this impact directly within the Structural Gravity Model framework as our approach is based on exports and so only reflects PE (partial equilibrium) effects – typically a full import/export CGE model is needed to compute GDP impacts. It is vital therefore to stress that our estimates of *Brexit* effects on national income are upper bound values as the counterweighting effect of *Brexit* on Irish imports from the UK is not incorporated. Nonetheless we have applied tariffs and/or diverted market scenarios directly to the CN6/HS6 values to get PE adjusted goods export values. Based on these we can directly compute *Hard* and *Long Term Hard Brexit* effects and adjust national income accordingly. Moreover, to smooth out recent the economic fluctuation in national income slightly we base our measure on the average value of goods exports in 2015 and 2016 as it applies to the average value of GNI* in 2015 and 2016 (CSO 2016b). GNI* *(i.e.* Modified GNI) is defined as GNI less the effects of the profits of re-domiciled companies and the depreciation of intellectual property products and aircraft leasing companies. This is a new indicator of the level of activity in the 'real' Irish economy introduced by CSO in 2016 primarily for debt ratio analysis. We compute the actual change in goods exports under *Hard* and *Long Term Hard Brexit* and apply these to the current value of GNI* with results reported in Table 10. Assuming imports and other factors remain as is and their counter weighting effect is not taken into account, the figures in Table 10 show that under a *Hard Brexit* the adjusted value of GNI* is €180.2bn while under a *Long Term Hard Brexit* scenario the value is €171.8bn representing a 0.4% and 5.1% drop in GNI* respectively. Interestingly, the latter long term effect drop of 5.1% aligns with the severe shock estimate of -6% of the UK Treasury quoted by the Dept. of Finance (DoF 2016) as part of the Budget analysis for 2017.

In summary the so-called *Soft-Brexit* scenario has no real effect on trade values so unsurprisingly this should be Ireland's preferred policy option. Interestingly our results show the overall effect of a *Hard Brexit* on the value of Irish goods exports is small at -1.4%. From a national income perspective this represents a fall of 0.4% in GNI*. Of course these are average effects; based on the discussion following Table 6 worst case 2 SE effects could be 1.4 times these values, giving a 1.96% effect on goods exports value and a 0.56% impact on GNI* - shock values that the Irish economy should be able to absorb. Nonetheless, across the individual sectors the effect of a *Hard Brexit* will be felt most in the traditional sectors where Ireland is most exposed to trading with the UK. Here the drop in export values will generally be in the region of 10 to 12%. Looking at the long term effects of a *Hard Brexit* the impact on the 'real' Irish economy as a whole is substantial with GNI* falling by over 5%. Under a 2 SE worst case situation this effect could be to reduce by €12.9bn and GNI* by 7.14%. Of course the burden of this fall in national income will fall on the traditional sectors and here our results showed that export trade with the UK would all but disappear in sectors such as Food and Beverage and Textiles. Moreover, finding new markets for these goods in the EU after *Brexit* will require very substantial reductions in export producer prices. Nonetheless, perhaps the most worrying feature about the results of this analysis is the range in magnitude of the impacts. The percentage falls in goods trade value with the UK for example go from 0% under a *Soft Brexit* up to 95% in some sectors under a worst case *Long Term Hard Brexit* scenario. Unsurprisingly with such a wide range in outcome values it would seem largely impossible for policy makers to chart a future course that deals with all eventualities. However this is not the case. In fact the span of outcome values reported here suggest that a core principle of Ireland's policy approach must centre on ensuring the UK and our EU colleagues fully understand the magnitude of economic impact of *Brexit* on Ireland both in lost income in traditional business but more so in the huge number of firms likely to cease trading. This will not just impact Irish incomes but will foster social dislocation and disaffection especially in those regions of Ireland most dependent on goods trade with the UK. Moreover, it is clear from this analysis that Ireland cannot simply trade (with the EU for example) its way out of the long term effects of a *Hard Brexit*. Without crying wolf, this fact along with social and economic and wider political impacts must be communicated with as much effort as practicable by the Irish Government and its officials.

# 7. CONCLUSIONS

This paper has examined the effects of *Brexit* on merchandise goods exports using the Gravity Model. The latest gravity modelling methodology has been used throughout. Firstly this includes embedding our approach with the realm of the latest economic theory via the Structural Gravity Model of Anderson and Van Wincoop (2003). The key innovation in this model is the inclusion of multilateral resistance terms (MRTs) to account for remoteness effects. We analyse goods exports only and so the Structural Gravity Model in this case recovers PE (partial equilibrium) effects but does not recover CGE (conditional general equilibrium) effects. Nonetheless we have endeavoured to incorporate MRTs into our approach. Second we have used Poisson Pseudo Maximum Likelihood (PPML) an estimation method popularised by Santos–Silva & Teyrano (2006) to generate our model parameter estimates rather than the more conventional Ordinary Least Squares (OLS) method. PPML has been shown to produce less bias parameter estimates than OLS for the Gravity Model and as a consequence has begun to supplant OLS as the estimation method of choice. We also have tested the robustness of PPML against dispersion using the alternative Negative Binomial PML approach and found PPML to be more reliable. Meanwhile we initially compared OLS and PPML estimates for Irish goods exports data and found the distance parameter estimate in the Classical Gravity Model is close to -1 for OLS while it is about -0.7 for PPML. Interestingly our PPML estimate is consistent with the finding of Santos–Silva & Teyrano (2006) – we highlight this as a key novel finding of this study.

Another novel feature of this research is our primary data source is the Central Statistics Office's (CSO) unit level CN 8-digit (Common Nomenclature) export data covering the years 1994 to 2016. We use this data, PPML estimation and a Remoteness Gravity Model structure that accounts for multilateral resistance to trade as a platform to generate a picture of the trading environment experienced by Irish goods exporters. Using this approach as a baseline our key findings show that including multilateral resistances has a substantial impact on parameter estimates. Our key novel finding here is that trade costs measured by distance has a substantial effect on goods exports, specifically a 1% increase in the distance Irish goods exports travel will mean the value of exports will drop by 0.73%. This value while large is nonetheless lower (in absolute magnitude) than the value found in previous studies such as Lawless (2010) or Fitzsimons et. al. (1999). Moreover, this baseline analysis showed that agricultural exports shifted to France from GB could see goods export values fall by up to 44% over time.

We followed up the modelling of goods exports by considering three possible *Brexit* scenarios in the context of the Remoteness Gravity Model; these scenarios are *Soft-Brexit, Hard Brexit and a Long Term Substitution Brexit* respectively. Interestingly our *Soft Brexit* analysis produced parameter estimates identical to the earlier baseline estimates showing that the UK remaining within the customs union has no effect on goods exports – as expected exporting life goes on as normal save for a nominal change in destination identity. Accordingly it makes sense that this option should remain Ireland's preferred policy option. However, if a *Hard Brexit* prevails goods exports will be impacted with a 1.4% drop on average with indigenous sectors of the economy taking the largest hit in value terms at 9 to 12%. The overall economic impact in the worst case will be no more than a ½% reduction in national income. Going beyond this the long term *Long Term Hard Brexit* goods export switching scenario showed trade costs would inevitably increase and goods exports value decreasing by over €9bn on average and national income falling by over 5%, while in the worst case the impact could be up to €13bn per year. Once again the traditional sectors of the Irish economy bear the brunt of this impact. We also showed that there is a wide range of realistic outcomes from those which we have quantified as having no impact on goods export values up to those having a 95% impact in some sectors. Accordingly, to address these threats to Ireland's future economic well-being we suggest Ireland must a) negotiate to ensure the UK maintains a full customs union and *regulatory alignment* with the EU that results in a *Soft Brexit*; or failing that b) negotiate to ensure the UK maintains full a *regulatory alignment* with the EU in the traditional sectors of the Irish economy most exposed to a *Hard Brexit;* or failing that c) negotiate a form *external association* for Northern Ireland within the EU customs union with associated management costs borne mainly via an Ireland Common Trading Fund; or failing that d) accept a *Hard Brexit* but obtain graduated exports credits for Irish goods exported to the UK over a number of years as part of the so-called *Brexit Bill* – this transition arrangement will facilitate Irish firms redirecting their activity away from the UK to the EU and allow the UK time to source cheaper markets outside the EU. One or a mix of these efforts will be necessary in order to ensure the UK remains supplied with quality products while staving off business annihilation in Ireland. This of course will allow the indigenous Irish economy time adjust to *Brexit* and shift focus beyond Great Britain - ironically an end many so-called *Brexiteers* want for their country in their relations with the EU. In any event, a widening of Ireland's trading horizon to the EU and beyond in the longer term is an outcome that would surely align with the aspirations of the authors of *Economic Development* back in 1958 and indeed resonate with many Irish people living today.

# APPENDIX

Table 1A: Parameter Estimates by NACE Sector Group for Irish Exports using Basic PPML (excludes MRTs)

| | Basic Exports Gravity Model PPML Parameter Estimates | | | | | | | | |
|---|---|---|---|---|---|---|---|---|---|
| | Main Industry Sector | | | | | | | | |
| Variable Name | Agriculture, Forestry & Fishing | Mining & Quarrying | Food & Beverage | Textiles | Wood & Paper | Chemicals, Pharma & Rubber | Metals & Machinery | Other Products | All Sectors |
| Intercept | -1.570 | -8.568 | 1.342 | 2.656 | 1.865 | -36.239* | -7.846* | -4.527* | -13.250* |
| GDP | 0.474* | 0.934* | 0.637* | 0.576* | 0.699* | 2.141* | 1.062* | 0.454* | 1.233* |
| Distance (km) | -0.705* | -1.275* | -0.781* | -1.338* | -1.431* | -0.419* | -0.644* | 0.591* | -0.530* |
| Area ($km^2$) | 0.224* | 0.327* | -0.081* | 0.016 | -0.037 | -0.095* | -0.222* | 0.087 | -0.098* |
| Population | 0.074 | -1.071* | 0.057 | 0.154 | -0.255* | -1.645* | -0.092* | -0.161* | -0.668* |
| Common Religion (Proportion Christian) | 0.152 | -0.19* | 0.024 | -0.081* | 0.022 | 0.418* | -0.001 | 0.23* | 0.187* |
| Great Britain | 2.066* | -0.252 | 0.145 | 0.074 | 0.177 | -0.088 | -0.855* | 4.141* | 0.045 |
| Northern Ireland | 2.371* | -2.807* | -0.494* | -0.386 | -1.537* | -2.146* | -1.483* | 3.204* | -1.373* |
| GATT/WTO member | -0.918 | 0.091 | -0.588* | 1.164* | -0.767* | 1.540* | 0.305 | -2.255* | -0.391 |
| English is official or primary language | -0.640* | 1.544* | -0.06 | 0.750* | 1.158* | -0.003 | 0.271 | 1.263* | 0.398* |
| EU member | 0.963* | -0.049 | 0.564* | -0.196 | 0.061 | -0.244 | 0.141 | 0.042 | -0.304* |
| Euro Currency | 0.571* | 1.770* | 0.102 | 0.538* | 0.404* | 0.548* | 0.256 | 2.841* | 0.612* |
| Common Legal System before Independence | 0.717 | -0.520 | 1.050* | -0.235 | 0.438* | 1.084* | 0.87* | -0.812* | 0.582* |
| Year | 0.010 | -0.156 | -0.063 | -0.474* | -0.059 | -0.125 | -0.261* | -0.525* | -0.141* |

\* denotes significance at 1% level

Table 2A: *Regulatory Alignment* Parameter Estimates by NACE Grouping for Irish Exports using PPML with MRTs

| | Remoteness Gravity Model | | | | | | | | |
|---|---|---|---|---|---|---|---|---|---|
| | Main Industry Sector | | | | | | | | |
| Variable Name | Agriculture, Forestry & Fishing | Mining & Quarrying | Food & Beverage | Textiles | Wood & Paper | Chemicals, Pharma & Rubber | Metals & Machinery | Other Products | All Sectors |
| Intercept | -4.72 | 4.323 | -5.945* | 30.917* | 6.846* | -6.053* | -19.818* | 31.521* | -11.309* |
| GDP | 0.433 | 0.658* | 0.592* | 0.207 | 0.569* | 1.265* | 1.19* | 0.058 | 1.066* |
| Distance (km) | -0.941* | -0.765 | -0.091 | -1.453* | -0.478* | -1.822* | -0.703* | 1.079* | -0.731* |
| Area ($km^2$) | 0.271* | 0.892* | 0.165* | -0.203* | 0.008 | -0.31* | -0.078* | -0.091 | -0.009 |
| Population | 0.516 | -1.239* | -0.098* | 0.555* | -0.387* | 0.183* | -0.151* | -0.066 | -0.341* |
| Common Religion (Proportion Christian) | 0.064 | -0.064 | 0.018 | -0.175* | -0.056* | 0.111* | -0.015 | 0.071 | 0.197* |
| Great Britain | 2.92* | 2.745 | 3.03* | -0.602 | 2.594* | -4.471* | -0.796* | 5.012* | -0.849* |
| Northern Ireland | 4.728* | 0.641 | 3.507* | -1.551 | 1.567* | -5.283* | -0.781* | 3.374* | -1.722* |
| GATT/WTO member | 1.180 | 0.583 | -0.561* | 1.144* | -0.535* | 0.068 | -0.025 | -1.819* | -1.047* |
| English is official or primary language | -1.506* | 0.263 | -0.344* | 0.345 | 0.875* | 0.407 | 1.623* | 0.836* | 1.068* |
| EU member | 0.974* | 0.019 | 0.979* | 0.101 | -0.156 | -0.501* | 0.452* | 0.362 | -0.191 |
| Euro Currency | 0.522* | 1.660* | -0.143 | 0.513* | 0.153 | -0.211 | -0.264* | 1.112* | -0.019 |
| Common Legal System before Independence | 0.796 | -1.133* | 0.661* | 0.198 | -0.027 | 1.132* | -0.393 | -1.13* | 0.073 |
| Export Remoteness (1%) (percentage of (5%) indicies significant) | 0  0 | 100  " | 0  " | 100  " | 100  " | 0  39 | 100  " | 100  " | 66  " |

\* denotes significance at 1% level